\documentclass[nojss]{jss}

\usepackage[utf8]{inputenc}

\author{
Paul-Christian Bürkner\\Aalto University, Department of Computer Science
}
\title{Bayesian Item Response Modeling in \proglang{R} with \pkg{brms} and
\proglang{Stan}}

\Plainauthor{Paul-Christian Bürkner}
\Plaintitle{Bayesian Item Response Modeling in R with brms and Stan}
\Shorttitle{Bayesian IRT Modeling with \pkg{brms}}

\Abstract{
Item Response Theory (IRT) is widely applied in the human sciences to
model persons' responses on a set of items measuring one or more latent
constructs. While several \proglang{R} packages have been developed that
implement IRT models, they tend to be restricted to respective
prespecified classes of models. Further, most implementations are
frequentist while the availability of Bayesian methods remains
comparably limited. I demonstrate how to use the \proglang{R} package
\pkg{brms} together with the probabilistic programming language
\proglang{Stan} to specify and fit a wide range of Bayesian IRT models
using flexible and intuitive multilevel formula syntax. Further, item
and person parameters can be related in both a linear or non-linear
manner. Various distributions for categorical, ordinal, and continuous
responses are supported. Users may even define their own custom response
distribution for use in the presented framework. Common IRT model
classes that can be specified natively in the presented framework
include 1PL and 2PL logistic models optionally also containing guessing
parameters, graded response and partial credit ordinal models, as well
as drift diffusion models of response times coupled with binary
decisions. Posterior distributions of item and person parameters can be
conveniently extracted and post-processed. Model fit can be evaluated
and compared using Bayes factors and efficient cross-validation
procedures.
}

\Plainkeywords{Item Response Theory, Bayesian Statistics, R, Stan, brms}

\Address{
    Paul-Christian Bürkner\\
  Aalto University, Department of Computer Science\\
  Konemiehentie 2, 02150 Espoo, Finland\\
  E-mail: \email{paul.buerkner@gmail.com}\\
  URL: \url{https://paul-buerkner.github.io/}\\~\\
  }

\usepackage{booktabs}
\usepackage{threeparttable}
\usepackage{booktabs}
\usepackage{longtable}
\usepackage{array}
\usepackage{multirow}
\usepackage{wrapfig}
\usepackage{float}
\usepackage{colortbl}
\usepackage{pdflscape}
\usepackage{tabu}
\usepackage{threeparttable}
\usepackage{threeparttablex}
\usepackage[normalem]{ulem}
\usepackage{makecell}
\usepackage{xcolor}

\usepackage{amsmath}

\begin{document}

\hypertarget{introduction}{%
\section{Introduction}\label{introduction}}

Item Response Theory (IRT) is widely applied in the human sciences to
model persons' responses on a set of items measuring one or more latent
constructs \citep[for a comprehensive introduction
see][]{lord2012, embretson2013, vanderlinden1997}. Due to its
flexibility compared to classical test theory, IRT provides the formal
statistical basis for most modern psychological measurement. The best
known IRT models are likely those for binary responses, which predict
the probability of a correct answer depending on the item's difficulty
and potentially other item properties as well as the participant's
latent ability. The scope of IRT models is however much wider than this,
and I will discuss several more interesting models in this paper.

Over the years, a multitude of software packages have been developed
that implement IRT models. To date, most free and open source software
in the field of IRT is written in the programming language \proglang{R}
\citep{R}, which has grown to become one of the primary languages for
statistical computing. Examples for widely applied and actively
maintained IRT specific \proglang{R} packages are \pkg{eRm} \citep{eRm},
\pkg{ltm} \citep{ltm}, \pkg{TAM} \citep{TAM}, \pkg{mirt} \citep{mirt},
\pkg{sirt} \citep{sirt}, and \pkg{psychotree}
\citep{psychotree1, psychotree2}. Each of them supports certain classes
of IRT models and related post-processing methods. Further, IRT models
may also be specified in general purpose multilevel or structural
equation modeling packages such as \pkg{lme4} \citep{lme4}, \pkg{lavaan}
\citep{lavaan}, \pkg{blavaan} \citep{blavaan}, or \pkg{MCMCglmm}
\citep{MCMCglmm}. I will provide a review and comparison of these
package later on in Section \ref{comparison}.

In this paper, I present a Bayesian IRT framework based on the
\proglang{R} package \pkg{brms} \citep{brms1, brms2} and the
probabilistic programming language \proglang{Stan}
\citep{carpenter2017}. The proposed framework is quite extensive both in
the models that can be specified and in the supported post-processing
methods. Users can choose from over 40 built-in response distributions,
which not only include standard IRT models such as binary, categorical,
or ordinal models, but also models for count data, response times or
proportions, to name only a few available options. Users may also write
their own custom response distributions not natively supported by
\pkg{brms} for application in the proposed framework. The non-linear
multilevel formula syntax of \pkg{brms} allows for a flexible yet
concise specification of multidimensional IRT models, with an arbitrary
number of person or item covariates and multilevel structure if
required. Prior knowledge can be included in the form prior
distributions, which constitute an essential part of every Bayesian
model. Estimation is performed in \proglang{Stan} using MCMC sampling
via adaptive Hamiltonian Monte Carlo \citep{hoffman2014, stanM2019}, an
efficient and stable algorithm that works well in high dimensional,
highly correlated parameter spaces. The proposed framework is designed
to help the applied researcher who wishes to analyze their IRT data by
means of one single package. However, it arguably requires more work
from the user at the start to familiarize themselves with the modeling
syntax and post-processing options and probably has a much steeper
learning curve than more specialized IRT packages. In addition,
\pkg{brms} also provides an opportunity for more methodologically
interested researchers who strive to develop new IRT models or model
variants. It could be a powerful and convenient tool to implement them
in a Bayesian context; with readily available data pre-processing and
model post-processing options, including, but not limited to,
summarizing and plotting parameters, computing and checking posterior
predictions, as well as performing model comparisons. These features
enable rapid model prototyping and testing, even if the final product is
eventually written in \proglang{Stan} itself or in another probabilistic
programming language.

This paper has three central purposes. First, it provides a thorough
conceptual introduction to the proposed Bayesian IRT framework. Second,
it demonstrates how this framework is implemented in statistical
software. Third, based on several hands-on examples, it explains how the
software can be used in practice to solve real-world questions. On the
conceptual side, in Section \ref{model}, I substantially extend the work
of \citet{deboeck2011}, who initially opened up the road for the
estimation of IRT models via multilevel models. However, they only
considered generalized linear multilevel models and specifically
focussed on binary data. I extend their framework in various directions,
most notably to (a) a much larger number of response distributions, (b)
non-linear IRT models, which do not make the assumption of the predictor
term being of a (generalized) linear form, and (c) distributional IRT
models, in which not only the main location parameter of the response
distribution but also all other parameters may depend on item and person
properties. On the software side, in Section \ref{brms} and
\ref{estimation}, I introduce several new features in \pkg{brms} that
have been implemented after the publication of its second paper
\citep{brms2} to both support the presented framework in its entirety
and provide several more specific features designed to make important
IRT model classes possible within the framework. These features include
the full integration of non-linear and distributional parameter
predictions via a nested non-linear formula syntax, the implementation
of several distributions designed for response times data, extentions of
distributions for ordinal data, for example for the purpose of modeling
discrimination parameters, and the ability to fix parameters to
arbitrary values. To help users applying the proposed framework and
related software in practice, several hands-on examples are discussed in
detail in Section \ref{examples}. I provide a comparison of IRT
supporting \proglang{R} packages in Section \ref{comparison} and end
with a conclusion in Section \ref{conclusion}. All materials related to
this paper are hosted on GitHub
(\url{https://github.com/paul-buerkner/Bayesian-IRT-paper}). To run all
code and examples shown in this paper, \pkg{brms} version 2.11.5 or
higher is required.

\hypertarget{model}{%
\section{Model description}\label{model}}

The core of models implemented in \pkg{brms} is the prediction of the
response \(y\) through predicting all \(K\) parameters \(\psi_k\) of the
response distribution \(D\). We write

\[
y_n \sim D(\psi_{1n}, \psi_{2n}, \ldots, \psi_{Kn})
\]

to stress the dependency on the \(n\textsuperscript{th}\) observation.
In most \proglang{R} packages, the response distribution is called the
model \code{family} and I adopt this term in \pkg{brms}. Writing down
the model per observation \(n\) implies that we have to think of the
data in long rather than in wide format. That is, reponses to different
items go in the same column of the data set rather than in different
columns. The long format works well in combination with multilevel
formula syntax and is arguably also more favourable from a programatical
perspective \citep[e.g., see][]{wickham2016}.

\hypertarget{respdists}{%
\subsection{Response distributions}\label{respdists}}

The response format of the items will critically determine which
distribution is appropriate to model individuals' responses on the
items. The possibility of using a wide range of response distributions
within the same framework and estimating all of them using the same
general-purpose algorithms is an important advantage of Bayesian
statistics. \pkg{brms} heavily exploits this advantage by offering a
multitude of response distributions and even allowing the user to add
their own. In this section, I will briefly review some common response
distributions in IRT that are natively supported in the proposed
framework.

If the response \(y\) is a binary success (1) vs.~failure (0) indicator,
the canonical family is the \emph{Bernoulli} distribution with density

\[
y \sim \text{Bernoulli}(\psi) = \psi^y (1-\psi)^{1-y},
\] where \(\psi \in [0, 1]\) can be interpreted as the success
probability. Common IRT models that can be built on top of the Bernoulli
distribution are the 1, 2, and 3 parameter logistic models \citep[1PL,
2PL, and 3PL models;][]{agresti2010}, which I will discuss in more
detail in Sections \ref{predDP} and \ref{binary}.

If \(y\) consitutes a categorical response with \(C > 1\) unordered
categories, the \emph{categorical} distribution is appropriate
\citep{agresti2010}. It has the density

\[
y \sim \text{categorical}(\psi_1, \ldots, \psi_C) = \prod_{c = 1}^C \psi_c^{I_c(y)}
\]

with cateory probabilities \(P(y = c) = \psi_c > 0\) and
\(\sum_{c=1}^C \psi_c = 1\) where \(I_c(y)\) is the indicator function
which evaluates to \(1\) if \(y = k\) and to \(0\) otherwise. For
\(C = 2\), the categorical distribution is equivalent to the Bernoulli
distribution.

If \(y\) is an ordinal categorical response with \(C\) ordered
categories, multiple possible response distributions are plausible
\citep{agresti2010, buerkner2019}. They are all built on top of the
categorical distribution but differ in how they define the category
probabilities \(P(y = c)\). The two most commonly applied ordinal
families in IRT are the \emph{cumulative model} and the \emph{adjacent
category model}. The cumulative model assumes

\[
P(y = c) = F(\tau_c - \psi) - F(\tau_{c-1} - \psi)
\] where \(F\) is the cumulative distribution function (CDF) of a
continuous unbounded distribution and \(\tau\) is a vector of \(C-1\)
ordered thresholds. If \(F\) is the standard logistic distribution, the
resuling IRT model is called \emph{graded response model}
\citep[GRM;][]{samejima1997}. Alternatively, one can use the adjacent
category model, which, when combined with the logistic distribution,
becomes the \emph{partial credit model} \citep[PCM;][]{rasch1961}. It
assumes

\[
P(y = c) = \frac{\exp \left(\sum_{j=1}^{c-1} (\psi -\tau_{j}) \right)}
  {\sum_{r=1}^{C} \exp\left(\sum_{j=1}^{r-1} (\psi -\tau_{j}) \right)}
\]

with threshold vector \(\tau\) whose element do not necessarily need to
be ordered \citep{adams2012}. The PCM is a widely used ordinal model in
IRT. One of its extentions, the generalized PCM (GPCM), even finds
application in various large scale assessment studies such as PISA
\citep{oecd2017}. I will provide hands-on examples of ordinal IRT models
in Section \ref{ordinal}.

If \(y\) consitutes a count variable without a natural upper bound (or
an upper bound that is practically not reachable, for instance in
dedicated speed tests), the \emph{Poisson} distribution with density

\[
y \sim \text{Poisson}(\psi) = \frac{\psi^y \exp(-\psi)}{y!},
\]

or one of its various generalizations \citep[e.g., see][]{shmueli2005},
may be an appropriate choice. In IRT, this leads to what is known as the
\emph{Rasch-Poisson-Counts model} \citep[RPCM;][]{rasch1960}.

When items consist of a comparative judgement between \(C\) categorical
alternatives on a continuous bounded scale, obtained responses are in a
``proportion-of-total'' (compositional) format \citep{hijazi2009}. That
is, for each response category \(c\), \(y_c \in [0, 1]\) is the
proportion of the total points that was assigend to that category so
that \(\sum_{c=1}^C y_c = 1\). If \(C = 2\), the response \(y = y_1\) on
the first category can be modeled as \emph{beta} distributed (as
\(y_2 = 1 - y_1\) is redundant). The mean-precision parameterization of
the beta distribution has density

\[
y \sim \text{Beta}(\psi_1 = \mu, \psi_2 = \phi) = 
\frac{y^{\mu \phi - 1} (1-y)^{(1-\mu) \phi-1}}{B(\mu \phi, (1-\mu) \phi)},
\]

where \(B\) is the beta function. A multivariate generalization of the
Beta family is the \emph{Dirichlet} family, which can be used for
compositional scores of more than two response categories
\citep{hijazi2009}. On the full response vector \(y = (y_1, ..., y_C)\)
it has density

\[
y \sim \text{Dirichlet}(\psi_1, \ldots \psi_C, \psi_{C+1} = \phi) = 
 \frac{1}{B((\psi_{1}, \ldots, \psi_{K}) \phi)} 
 \prod_{k=1}^K y_{k}^{\psi_{k} \phi - 1}.
\]

Another important class of IRT models deals with response/reaction
times, which tend to vary over items and persons in at least three ways:
mean, variation, and right skewness of the responses. Accordingly,
sufficiently flexible response distributions on reaction times are
likely to require three parameters in order to capture these aspects.
Two commonly applied 3-parameter distributions are the
exponentially-modified Gaussian (exgaussian) distribution and the
shifted lognormal distribution \citep{heathcote1991, wagenmakers2007}.
Their densities are a little bit more involved and so I do not display
them here, but they can be found for instance in \citet{wagenmakers2007}
or when typing \texttt{vignette("brms\_families")} in \proglang{R}. With
the exgaussian distribution, we can directly parameterize the mean which
simplifies interpreation of model parameters, at the expense of having a
theoretically less justified model \citep{heathcote1991}. I will provide
a practical example of analyzing response times in an IRT context in
Section \ref{response-times}.

Going one step further, it is often favorable to model persons'
responses together with the corresponding response times in a joint
process model. This not only implies a more appropriate generative model
for the data but may also foster theoretical understanding of the
underlying processes \citep{ratcliff1978, vandermaas2011}. One of these
joint models, which can handle binary decisions together with their
response times, is the Wiener drift diffusion model
\citep{ratcliff1978, vandermaas2011}. Its parameters have meaning in the
context of a cognitive decision process described as a Wiener diffusion
process with a drift towards one or the other binary choice alternative.
The parameters of the four parameter drift diffusion model implemented
in the presented framework are (1) the drift rate that describes a
person's tendency towards one or the other two alternatives, (2) the
boundary separation that describes how much evidence needs to be
accumulated until a decision is made, (3) the non-decision times that
describes the time spent at processing the items and executing a motor
response (i.e., everything non-decision related), and (4) the initial
bias that describes persons tendency towards one of the two alternatives
independent of the item properties. In IRT applications, it is common to
fix the initial bias to \(0.5\), that is, to assume no initial bias
towards one of the two alternatives \citep{diffIRT}, which results in
the three-parameter drift diffusion model. A more detailed discussion of
the drift diffusion models is beyond the scope of the present paper, but
can be found elsewhere \citep{ratcliff1978, vandermaas2011, diffIRT}. I
will provide a practical example of fitting drift diffusion models to
IRT data in Section \ref{response-times}.

\hypertarget{predDP}{%
\subsection{Predicting distributional parameters}\label{predDP}}

In the context of IRT, every distributional parameter \(\psi_k\) can be
written as a function \(\psi_{kn} = f_k(\theta_{kp_n}, \xi_{ki_n})\) of
person parameters \(\theta_{k}\) and item parameters \(\xi_{k}\), where
\(p_n\) and \(i_n\) indicate the person and item, respectively, to which
the \(n\textsuperscript{th}\) observation
belongs\footnote{A parameter may also be assumed constant across observations
and thus be independent of person and item parameters.}. In a regression
context, such models are often referred to as distributional regression
models or as regression models of location, scale, and shape
\citep{rigby2005} to stress the fact that all parameters of the
distribution can be predicted, not just a single parameter -- usually
the mean of the distribution or some other measure of central tendency.

In addition to the response distribution itself, the exact form of the
equations \(\psi = f(\theta_p, \xi_i)\) (suppressing the indices \(k\)
and \(n\) for simplicty) will critically define the meaning of the
person and item parameters as well as the complexity of the model in
general. In a linear model, \(f\) is the identity function and the
relation between \(\theta_p\) and \(\xi_i\) is linear and additive so
that \(\psi = \theta_p + \xi_i\). Unfortunately, such a model will not
yield the desired results if \(\psi\) has natural range restrictions.
For instance, if the response \(y\) is a binary success (1) vs.~failure
(0) indicator, and we use the Bernoulli response distribution, \(\psi\)
can be interpreted as the success probability, which, by definition,
must lie within the interval \([0, 1]\). However, a linear model
\(\psi = \theta_p + \xi_i\) may yield any real value and so is invalid
when predicting probabilities. The solution for this problem is to use a
non-linear function \(f\) appropriate to the scale of the predicted
parameter \(\psi\). This results in what is known as a \emph{generalized
linear model} (GLM). That is, the predictor term
\(\eta = \theta_p + \xi_i\) is still linear but transformed, as a whole,
by a non-linear function \(f\), which is commonly called `response
function'. For Bernoulli distributions, we can canonically use the
logistic response function

\[
f(\eta) = \text{logistic}(\eta) = \frac{\exp(\eta)}{1 + \exp(\eta)},
\]

which yields values \(f(\eta) \in [0, 1]\) for any real value \(\eta\).
As a result, we could write down the model of \(\psi\) as

\[
\psi = \frac{\exp(\theta_p + \xi_i)}{1 + \exp(\theta_p + \xi_i)},
\]

which is known as the Rasch or 1PL model \citep{bond2013}. Under the
above model, we can interprete \(\theta_p\) as the ability of person
\(p\) in the sense that higher values of \(\theta_p\) imply higher
success probabilities regardless of the administered item. Further, we
can interprete \(\xi_i\) as the easiness of item \(i\) as higher values
of \(\xi_i\) imply higher success probabilities regardless of the person
to which the item is administered. Note that most definitions of the
Rasch model instead use \(\theta_p - \xi_i\), in which case \(\xi_i\)
becomes the item difficulty rather than the easiness. Clearly, both
formulations are equivalent. In the present paper I generally use the
easiness formulation as it naturally fits into the regression framework
of \pkg{brms}.

In the context of IRT, GLMs already will carry us a long way, but at
some point, their flexibility reaches a halt. A typical example of such
a situation is when we stop assuming discriminations to be constant
across items; an assumption that will often be violated in real world
data \citep{andrich2004}. Instead, if we want to model varying item
discrimations \(\alpha_i\), the predictor term becomes

\[
\psi = f(\alpha_i (\theta_p + \xi_i)) = f(\alpha_i \theta_p + \alpha_i \xi_i).
\]

The argument to \(f\) no longer forms a linear predictor as we now
consider \emph{products} of parameters. In the context of logistic
models for dichotomous responses, we would refer to the varying
discrimination model as 2PL model \citep[e.g.,][]{andrich2004} or
generalized nonlinear mixed effects model \citep[e.g.,][]{deboeck2004}.
If persons have a non-zero probability \(\gamma_i\) of guessing the
right answer of item \(i\), independently of their abilities, this would
yield the 3PL model, in my notation written as

\[
\psi = f(\theta_{p}, \xi_{i}, \alpha_i, \gamma_i) = 
\gamma_i + (1 - \gamma_i) \, g(\alpha_i (\theta_p + \xi_i))
\]

with \(g\) being some function to transform real values onto the unit
interval (e.g., the logistic function). The complexity of such a
non-linear predictor may be arbitrarily increased in theory, but of
course needs to be carefully specified in order to yield an identifiable
and interpretable model \citep[see also][]{buerkner2020spm}. Further, in
the context of Bayesian IRT, prior distributions may additionally help
to identify the model (see Section \ref{priors} for more details on
priors).

\hypertarget{item-and-person-covariates}{%
\subsection{Item and Person
Covariates}\label{item-and-person-covariates}}

A lot of research questions in the context of IRT do not simply require
estimating person and item parameters but rather estimating the effects
of person or item \emph{covariates} \citep{deboeck2011}, that is
variables that vary across persons and/or items. \citet{deboeck2011}
differentiate covariates by their mode (person, item, or both) and the
origin of the covariate as either internal (stems from item responses)
or external (independent of the item responses). For instance, persons'
age would be considered an external person covariate as it varies over
persons but not over items and does not change its value according to
item responses. Item type (e.g., figural, numeric, or verbal in case of
typical intelligence test items) would be considered an external item
covariate, while the number of previous items solved by a specific
person at the time of administering a specific item would be an internal
person-by-item covariate.

Regardless of the specific nature of the covariates, we may add them to
any linear predictor term \(\eta\) in the model so that it no longer
only depends on individual person and item parameters, but also on a set
of \(J\) covariates \(x_j\):

\[
\eta_{pi} = \theta_p + \xi_i + \sum_{j=1}^J b_j x_{jpi}
\]

In the equation above, \(x_{jpi}\) is the value of the \(j\)th predictor
for person \(p\) and item \(i\). Of course, a person covariate is
constant across items and an item covariate is constant across persons.
I still index all covariates by both persons and items, though, to
shorten the notation without loss of generality.

A further differentiation of covariates may be made by considering over
what mode (person, items, or both) the covariate effects are allowed to
vary (i.e., interact with) in the model. For example, a persons' age
varies between but not within persons, which implies that the
\emph{effect} of age may only vary across items. Conversely, the effect
of an item covariate may only vary across persons as it is constant
within each item. Extending the above notation for covariates, the
regression coefficients \(b_j\) would then receive additional indices
\(p\) or \(i\) (i.e., \(b_{jp}\) or \(b_{ji}\)) depending on whether the
effect of the covariate is expected to vary over persons or items.

Depending on the nature of the covariates and over which mode their
effects are assumed to vary, the full model may not be identified or at
least hard to estimate and interprete. Thus, careful specification of
covariates is critical to obtain sensible results. \citet{deboeck2011}
provide a thoughtful and thorough discussion of the use of covariates in
IRT models and I do not want to reiterate every detail, but simply note
that all kinds of covariate models discussed in their paper may be
specified in the here presented framework using the same formula syntax.
Some examples for covariate analysis are provided in Section
\ref{binary}.

\hypertarget{differential-item-functioning}{%
\subsection{Differential item
functioning}\label{differential-item-functioning}}

For psychometric tests, it is essential to investigate differential item
functioning \citep[DIF;][]{holland1993, osterlind2009}. Items showing
DIF have different properties for persons belonging to different groups
even if the persons have the same ability. Such items may reduce test
validity as they hinder measurement equivalence and may lead to bias in
the latent trait estimates \citep[e.g.,][]{millsap1993, holland1993}.
However, simply excluding all items showing DIF can be equally
problematic, as, for example, the resulting subtest may contain a too
homogeneous set of items that fail to cover all relevant aspects of the
constructs the test claims to measure \citep[e.g.,][]{osterlind2009}.
Thus, removing DIF items can also be a threat to validity. As
\citet{osterlind2009} put it: ``We emphazise that the DIF phenomenon is
complex, and one should have a sophisticated -- even wise --
understanding before one approaches questions of how to address it in a
given test situation.'' It is beyond the scope of this paper to provide
such wisdom, but what we can do is apply IRT models to detect (certain
kinds of) DIF in the first place.

In turns out that DIF analysis can be performed by including and
analyzing specific person-by-item covariates. That is, we need (a) a set
of person covariates, such as age, gender, or sociodemographic status,
defining groups of people for which DIF should be tested, and (b) a set
of item properties over which to expect DIF; or simply an item indicator
if we want to test DIF over each individual item. The interactions
between the selected person covariates/groups and item
covariates/indicators then form the DIF variables. If, after inclusion
of the involved person covariates' main effects, these interactions show
predictive value for the response, this indicates DIF. I show a hands-on
example of covariate-based DIF analysis in Section \ref{binary}. A more
detailed and technical discussion about this approach is provided in
\citet{deboeck2011}.

A common difficulty in performing DIF analysis is the sheer amount of
potential DIF variables that can be investigated; much more than we can
usually check in practice let alone include in a single model. A
powerful option is to use variable selection techniques to select the
relevent DIF variables from a larger set of potentially relevant
variables, for instance, using lasso methods \citep{schauberger2019}. In
Bayesian statistics, variable selection can be implemented in the form
of two-stage procedures. First, an all encompassing model including all
covariates is fitted while imposing strongly regularizing/sparsifying
priors such as the (regularized) horseshoe prior
\citep{carvalho2010, piironen2017} to keep the model estimable and
prevent overfitting (see Section \ref{priors} for a formal introduction
to prior distributions in an IRT context). In theory, selection can be
done already after this first step but the results tend to be suboptimal
in a lot of cases \citep{piironen2017comparison}. For variable selection
to work well, an explicit variable selection procedure, such as
projective predictions, has to be applied in a second step, in which the
full model serves as a reference \citep{piironen2017comparison}. While
the first step is already possible for complex multilevel models, the
second step still needs more research, which we plan on doing in the
upcoming years.

\hypertarget{priors}{%
\subsection{Prior distributions of person and item
parameters}\label{priors}}

In Bayesian statistics, we are interested in the posterior distribution
\(p(\theta, \xi | y)\) of the person and item parameters given the
data\footnote{In IRT covariate models, the posterior distribution also includes
the covariates' coefficients and all hyperparameters, but I keep this implicit
in the equations to simplify the notation.}. The posterior distribution
is computed as

\[
p(\theta, \xi | y) = \frac{p(y | \theta, \xi) \, p(\theta, \xi)}{p(y)}.
\]

In the above equation \(p(y | \theta, \xi)\) is the likelihood,
\(p(\theta, \xi)\) is the prior distribution and \(p(y)\) is the
marginal likelihood. The likelihood \(p(y | \theta, \xi)\) is the
distribution of the data given the parameters and thus relates to the
data to the parameters. We may also describe the likelihood as the
combination of response distribution and predictor terms discussed
above. The prior distribution \(p(\theta, \xi)\) describes the
uncertainty in the person and item parameters before having seen the
data. It thus allows to explicitely incorporate prior knowledge into the
model. In practice, we will factorize the joint prior \(p(\theta, \xi)\)
into the product of \(p(\theta)\) and \(p(\xi)\) so that we can specify
priors on person and items parameters independently. The marginal
likelihood \(p(y)\) serves as a normalizing constant so that the
posterior is an actual probability distribution. Except in the context
of specific methods (i.e., Bayes factors), \(p(y)\) is rarely of direct
interest.

In frequentist statistics, parameter estimates are usually obtained by
finding those parameter values that maximise the likelihood. In
contrast, Bayesian statistics aims to estimate the full (joint)
posterior distribution of the parameters. This is not only fully
consistent with probability theory, but also much more informative than
a single point estimate (and an approximate measure of uncertainty
commonly known as `standard error').

Obtaining the posterior distribution analytically is only possible in
certain cases of carefully chosen combinations of prior and likelihood,
which may considerably limit modeling flexibilty but yield a
computational advantage. However, with the increased power of today's
computers, Markov-Chain Monte-Carlo (MCMC) sampling methods constitute a
powerful and feasible alternative to obtaining posterior distributions
for complex models in which the majority of modeling decisions is made
based on theoretical and not computational grounds. Despite all the
computing power, these sampling algorithms are computationally very
intensive and thus fitting models using full Bayesian inference is
usually much slower than in point estimation techniques. However,
advantages of Bayesian inference -- such as greater modeling
flexibility, prior distributions, and more informative results -- are
often worth the increased computational cost \citep{gelman2013}. For a
comprehensive introduction to Bayesian IRT modeling see, for example,
\citet{fox2010}, \citet{levy2017}, as well as \citet{rupp2004}.

I will start explaining important aspects concerning the choice of
priors for item parameters. A key decision when setting up an IRT model
is whether we want item parameters to share a common hierarchical prior
or if we want to use non-hierarchical and thus independent priors on
each parameter. In IRT, non-hierarchical priors are currently applied
more commonly \citep[e.g.,][]{fox2010, levy2017}. In the
non-hierarchical case, we would choose a prior and then fix its
hyperparameters according to our understanding of the scale and prior
knowledge about the parameter(s) to be estimated. To make a concrete
example, we can assume a normal distribution with mean \(0\) and
standard deviation \(3\) for the item easiness parameters of a Rasch
model:

\[
\xi_i \sim \text{Normal}(0, 3)
\]

By definition of the normal distribution, we thus assume a-priori, that
with \(68\%\) probability easiness parameters lie within \([-3, 3]\) and
that with \(97.5\%\) probability easiness parameters lie within
\([-6, 6]\) on the logit scale. Given the scale of the logistic response
function, this prior can be considered weakly informative. That is, it
restricts the parameters to a reasonable range of values without
strongly influencing the obtained posterior distribution. Of course, we
don't need to restrict ourselves to normal distributions. Other prior
distributions, such as a student-t distribution are possible as well,
although assuming a normal distribution is arguably a good default
choice \citep[see also][]{mcelreath2017}.

A fundamentally different class of priors arises when assuming the item
parameters to have the same underlying hierarchical prior distribution
with shared hyperparameters. Most commonly, a centered normal
distribution is used so that

\[
\xi_i \sim \text{Normal}(0, \sigma_\xi)
\]

for all \(\xi_i\), which share a common standard deviation
\(\sigma_\xi\). The latter is estimated as part of the model as well.
Such a prior implies that parameters are shrunken somewhat towards their
joint mean, a phenomenon also known as \emph{partial pooling}
\citep{gelmanMLM2006}. Partial pooling makes parameter estimates more
robust as well as less influenced by extreme patterns and noise in the
data \citep{gelmanMLM2006}.

The above model formulation implies that multiple item parameters
belonging to the same item but estimated for different distributional
parameters are assumed independent of each other. This turns out to be
an unnecessarily restrictive assumption in many applications. For
example, in a 2PL model, every item has both a difficulty/easiness and a
discrimination parameter, which may correlate with each other over the
set of items. At the very least, we cannot be sure of their independence
a-priori. Thus, accounting for their possible dependence appears to be
the generally safer choice. Statistically, correlated item parameters
are modeled via a hierarchical multivariate normal distribution in the
form of

\[
(\xi_{1i}, \ldots, \xi_{Ki}) \sim \text{Multinormal}(0, \mathbf{\Sigma_\xi})
\]

where \(\xi_{ki}\) is the item parameter of item \(i\) used in the
prediction of the distributional parameter \(\psi_k\) and
\(\mathbf{\Sigma_\xi}\) is the covariance matrix determining both the
scale and the dependence structure of the item parameters. A covariance
matrix tends to be relatively hard to interpret. Accordingly it is
usually advantageous to decompose the covariance matrix into a
correlation matrix capturing the dependence structure and a vector of
standard deviations capturing the scales of the person parameters:

\[
\mathbf{\Sigma_\xi} = \mathbf{D}(\sigma_{\xi 1}, \ldots, \sigma_{\xi K}) \, \mathbf{\Omega_\xi} \, \mathbf{D}(\sigma_{\xi 1}, \ldots, \sigma_{\xi K})
\]

In the above equation, \(\mathbf{\Omega_\xi}\) denotes the correlation
matrix and \(\mathbf{D}(\sigma_{\xi 1}, \ldots, \sigma_{\xi K})\)
denotes the diagonal matrix with standard deviations \(\sigma_{\xi k}\)
on the diagonal. Currently, \pkg{brms} officially supports only normal
hierarchical priors but other options, such a (multivariate) student-t
priors may be supported in the future as well.

A decision between hierarchical and non-hierarchical priors is not
always easy. If in doubt, one can try out both kinds of priors and
investigate whether they make a relevant difference since both are
supported in \pkg{brms}. Personally, I prefer hierarchical priors as
they imply some data-driven shrinkage due to their scale being learned
by the model on the fly. Also, they naturally allow item parameters to
share information across parameter classes via the correlation matrix
\(\mathbf{\Omega_\xi}\). Hierarchical item priors are currently a
somewhat non-standard approach in IRT. One reason might be that,
historically, a commonly used logic for using hierarchical distributions
was the assumption of an underlying population -- especially from a
frequentist viewpoint. In cases where we just have a fixed set of items
\citep[as opposed to some rule-based automatic generation of items;
e.g.,][]{gierl2012}, the assumption of an underlying population indeed
seems a bit off. However, I argue that for the application of
hierarchical priors, we actually do not need this population assumption,
especially not from a Bayesian perspective where uncertainty is
represented via probability distributions anyway. Rather, the decision
on whether or not to apply hierarchical priors should be based on
whether or not partial pooling of parameters is desired in a given
context.

In contrast to priors on item parameters, priors on person parameters
are almost always hierarchical in IRT
\citep[e.g.,][]{deboeck2011, fox2010, levy2017}. I will follow this
approach throughout this paper although a no pooling approach could be
adopted in \pkg{brms} as well. The procedure is similar to that of
hierarchical item priors. If we just have a single parameter
\(\theta_p\) per person \(p\) (or want to model multiple person
parameters as uncorrelated), we apply a hierarchical univariate prior of
the form

\[
\theta_p \sim \text{Normal}(0, \sigma_\theta)
\]

with a common standard deviation parameter \(\sigma_\theta\). For
multiple (correlated) person parameters we use an hierarchical
multivariate normal prior

\[
(\theta_{1p}, \ldots, \theta_{Kp}) \sim \text{Multinormal}(0, \mathbf{\Sigma_\theta})
\] with a common covariance matrix \(\mathbf{\Sigma_\theta}\), which may
again be decomposed into a vector of standard deviations and a
correlation matrix.

If we decide to partially pool both person and item parameters, we have
to amend the model slightly by adding an overall intercept parameter
\(b_0\) to the linear predictor, which then becomes
\(b_0 + \theta_p + \xi_i\). We do this in order to catch average
deviations from zero, which would otherwise no longer be appropriately
modeled as both person and item parameters had been (soft) centered
around zero by the prior. Such a formulation of IRT models via partially
pooled person and/or item parameters moves them into the framework of
\emph{generalized linear multilevel models} (GLMMs) and allows
corresponding GLMM software to fit certain kinds of IRT models
\citep{deboeck2011}.

What remains to be specified are priors on the hyperparameters, that is,
on the standard deviations and correlation matrices, if present. In
short, for standard deviations, I recommend priors whose densities have
a mode at zero and fall off strictly monotonically for increasing
parameter values. Examples for such priors are half-normal or
half-cauchy priors. For correlation matrices, I recommend the LKJ prior
\citep{lewandowski2009}, with which we can assign equal density over the
space of valid correlation matrices if desired. More details on
hyperparameters in \pkg{brms} and \proglang{Stan} are provided in
\citet{brms1}, \citet{brms2}, and the Stan User's Manual
\citep{stanM2019}.

Lastly, I want to discuss priors on covariate effects. A special
complexity in that context is that the scale of the coefficients depends
not only on the (link-transformed) scale of the response variable but
also on the scale of the covariates themselves (and possibly also on the
dependency between covariates). Additionally, the choice of priors
depends on the goal we want to achieve by their means, for instance,
improving convergence, penalizing unrealisticly large values, or
covariate selection \citep[see also][]{gelman2017}. \pkg{brms} supports
several covariate priors, ranging from completely flat ``uninformative''
priors (the current default), over weakly-informative priors for mild
regularization and improving convergence to priors intended for variable
selection such as the horseshoe prior
\citep{carvalho2010, piironen2017}. In general, setting priors is an
active area of research and I hope that we can further improve our
understanding of and recommendations for priors in the future.

\hypertarget{identification}{%
\subsection{Model Identification}\label{identification}}

An important aspect of model building is model identification
\citep[e.g.,][]{vanderlinden1997}. From a frequentist perspective, we
call a model \emph{identified} (or \emph{strongly identified} to
distinquish it from other notions of identification) if equality of
likelihood implies equality of parameter values
\citep[e.g.,][]{sanmartin2015}:

\begin{equation}
\label{def_identification}
p(y | \theta_1, \xi_1) =  p(y | \theta_2, \xi_2)
\quad \Rightarrow \quad
(\theta_1, \xi_1) = (\theta_2, \xi_2)
\end{equation}

If the above implication does not hold, certain parameters need to be
fixed to ensure identification or other constraints be made
\citep[e.g.,][]{sanmartin2015, sanmartin2010, bollen2009}. An important
example for a non-identified models (in a frequentist sense) are binary
2PL models when the standard deviation of the person parameters is
allowed to vary freely. This is because the scale of the person
parameters' distribution is completely accounted for by the scale of the
discrimination parameters due to their multiplicative relationship. It
is thus common practice to fix \(\sigma_\theta\) to \(1\) which then
ensures identification of the model \citep[e.g.,][]{vanderlinden1997}.

By the rules of probability theory, the posterior distribution of a
model always exists if all prior distributions are proper (i.e.,
integrate to \(1\)) independently of whether or not the model is
identified in the strong sense. As a result, identification of a
Bayesian model is not as straight forwardly defined as for a frequentist
model \citetext{\citealp[see][ for a thorough discussion on notions of
Bayesian identifiability]{sanmartin2010}; \citealp[see
also][]{fox2010}; \citealp[and][ for discussions in the context of IRT
models]{levy2017}}. In this context, I use the informal term
\emph{weakly identified} to describe a model that has a sensible
posterior distribution for all quantities of interest. The term
\emph{sensible} can only be resolved in the context of a given model and
data \citep[see also][]{gelman2017}. We will usually not be able to
check weak identification directly, as the posterior distribution is not
analytical for most Bayesian IRT models. Hence, in practice, weak
identification can only be verified empirically for a given estimation
algorithm based on whether the algorithm converges to a sensible
posterior distribution. In this sense, identification and convergence of
the estimation algorithm cannot easily be distinquished from one
another. What we can do is to use (non-)convergence as a proxy for
(non-)identification in addition to using our subject matter knowledge
about the expected scale of parameters values. However, care must be
taken as non-convergence may have several reasons only one of which is
non-identification.

In the present paper, I will go the way of imposing constraints commonly
made to ensure strong identification, such as fixing certain parameters
to constants. This also renders the model results comparable to those
obtainable by other frameworks which do not have full access to priors
on all parameters as a way to (weakly) identify models.

\hypertarget{brms}{%
\section{Model specification in brms}\label{brms}}

In \pkg{brms}, specifying an IRT model is done mainly via three
arguments: \texttt{family}, \texttt{formula}, and \texttt{prior}. I will
explain each of them in detail in the following.

\hypertarget{family-prior}{%
\subsection{Specifying the family argument}\label{family-prior}}

The model \texttt{family} specifies the response distribution as well as
the response functions of the predicted distributional parameters.
Following the convention of GLM theory, I do not specify the response
function directly but rather its inverse, which is called the link
function\footnote{In my opinion, the
convention of specifying link functions instead of response functions is
unfortunate. I think it is more natural to transform linear predictors to the
scale of the parameter via the response function, rather than transforming the
parameter to the scale of the linear predictor.}. In \pkg{brms}, each
response distribution has a dedicated primary parameter \(\psi_1 = \mu\)
that usually describes the mean of the distribution or some other
measure of central tendency. This primary parameter is accompanied by a
corresponding link function, which, as explained above, ensures that
\(\mu\) is on the scale expected by the distribution. In the \pkg{brms}
framework, a \texttt{family} can be specified via

\begin{CodeChunk}

\begin{CodeInput}
R> family = brmsfamily(family = "<family>", link = "<link>")
\end{CodeInput}
\end{CodeChunk}

where \texttt{\textless{}family\textgreater{}} and
\texttt{\textless{}link\textgreater{}} have to be replaced by the names
of the desired response distribution and link function of \(\mu\),
respectively. For binary responses, we could naturally assume a
Bernoulli distribution and a \texttt{logit} function, which would then
be passed to \pkg{brms} via

\begin{CodeChunk}

\begin{CodeInput}
R> family = brmsfamily(family = "bernoulli", link = "logit")
\end{CodeInput}
\end{CodeChunk}

The Bernoulli distribution has no additional parameters other than
\(\mu\), but most other distributions do. Take, for instance, the normal
distribution, which has two parameters, the mean \(\mu\) and the
residual standard deviation \(\sigma\). The mean paramter \(\mu\) can
take on all real values and thus, using the identity link (i.e., no
transformation at all) is a viable solution. If we assumed \(\sigma\) to
be constant across observations, we would simply specify

\begin{CodeChunk}

\begin{CodeInput}
R> family = brmsfamily(family = "gaussian", link = "identity")
\end{CodeInput}
\end{CodeChunk}

If, however, we also modeled \(\sigma\) as depending on item and/or
person parameters, we would need to think of a link function for
\(\sigma\) as well. This is because \(\sigma\) is a standard deviation,
which, by definition, can only take on positive values. A natural choice
to restrict predictions to be positive is the log link function with the
corresponding exponential response function, which is used as the
default link for \(\sigma\). To make this choice explicit, we write

\begin{CodeChunk}

\begin{CodeInput}
R> family = brmsfamily(family = "gaussian", link = "identity",
R>                     link_sigma = "log")
\end{CodeInput}
\end{CodeChunk}

An overview of available families in \pkg{brms} together with their
distributional parameters and supported link functions is provided in
\texttt{?brmsfamily}. Details about the parameterization of each family
are given in \code{vignette("brms\_families")}. If the desired response
distribution is not available as a built-in family, users may specify
their own custom families for use in \pkg{brms}. Details on custom
families can be found by typing \code{vignette("brms\_customfamilies")}
in the console.

\hypertarget{formula-prior}{%
\subsection{Specifying the formula argument}\label{formula-prior}}

I will now discuss the \texttt{formula} argument of \pkg{brms}.
Throughout this paper, I will assume the response variable to be named
\texttt{y} and the person and item indicators to be named
\texttt{person} and \texttt{item}, respectively. Of course, these names
are arbitary and can be freely chosen by the user as long as the
corresponding variables appear in the data set. If we just predict the
main parameter \(\mu\) of the response distribution, we just need a
single \proglang{R} formula for the model specification. If we want to
apply partial pooling (i.e., hierarchical priors) to the person
parameters but not to the item parameters, we would write

\begin{CodeChunk}

\begin{CodeInput}
R> formula = y ~ 0 + item + (1 | person)
\end{CodeInput}
\end{CodeChunk}

Instead, if we wanted to partially pool both person and item parameters,
we would write

\begin{CodeChunk}

\begin{CodeInput}
R> formula = y ~ 1 + (1 | item) + (1 | person)
\end{CodeInput}
\end{CodeChunk}

Throughout this paper, I will model both person and item parameters via
partial pooling as I believe it to be the more robust approach, which
also scales better to more complex models \citep{gelmanMLM2006}. If
partial pooling of items is not desired, the expression
\texttt{1\ +\ (1\ \textbar{}\ item)} has to be replaced by
\texttt{0\ +\ item}.

In standard \proglang{R} formula syntax, from which \pkg{brms} formula
syntax inherits, covariates may be included in the model by adding their
names to the formula. For instance, if we wanted to model an overall
effect of a covariate \texttt{x}, we would write

\begin{CodeChunk}

\begin{CodeInput}
R> y ~ 1 + x + (1 | item) + (1 | person)
\end{CodeInput}
\end{CodeChunk}

Additionally, if we wanted the effect of \texttt{x} to vary over items,
we would write

\begin{CodeChunk}

\begin{CodeInput}
R> y ~ 1 + x + (1 + x | item) + (1 | person)
\end{CodeInput}
\end{CodeChunk}

Modeling covariate effects as varying over persons can be done
analogously. Interactions are specified via the \texttt{:} operator.
That is, for covariates \texttt{x1} and \texttt{x2} we add
\texttt{x1:x2} to the formula in order to model their interaction. We
may also use \texttt{x1\ *\ x2} as a convenient short form for
\texttt{x1\ +\ x2\ +\ x1:x2}. As the data is expected to be in long
format, the syntax for covariate effects is independent of the covariate
type, that is, whether it is person or item related.

In most basic IRT models, only the mean of the response distribution is
predicted while other distributional parameters, such as the residual
standard deviation of a normal distribution, are assumed constant across
all observations. Depending on the psychometric test, this may be too
restrictive an assumption as items and persons not only differ in the
mean response but also in other aspects, which are captured by
additional parameters. To predict multiple distributional parameters in
\pkg{brms}, we need to specify one formula per parameter as follows:

\begin{CodeChunk}

\begin{CodeInput}
R> formula = bf(
R>   y ~ 1 + (1 | item) + (1 | person),
R>   par2 ~ 1 + (1 | item) + (1 | person),
R>   par3 ~ 1 + (1 | item) + (1 | person),
R>   ...
R> )
\end{CodeInput}
\end{CodeChunk}

The function \code{bf} is a short form for \code{brmsformula}, which
helps to set up complex models in \pkg{brms}. In the specification
above, \code{par2} and \code{par3} are placeholders for the parameter
names, which are specific to each response distribution, for instance,
\code{sigma} in the case of the normal distribution. Covariates effects
on such parameters may be included in the same way as described before.

The model formulation shown above implies that person and item
parameters, respectively, imposed on different distributional parameters
are modeled as independent of each other. However, to allow the exchange
of information between parameters of the same person or item and improve
partial pooling across the whole model, it is beneficial to specify them
as correlated (see Section \ref{priors} for details). Taking the 2PL
model as an example, we have two parameters per item, one difficulty and
one discrimination parameter. If, say, more difficult items also have
higher discrimination, we would like to use this information in the
model to improve the estimation of both difficulties and discriminations
even if we are not interested in the correlation itself \citep[e.g.,
see][ for a general discussion of correlations between varying
coefficients in multilevel models]{gelmanMLM2006}.

The syntactical solution to model these correlations implemented in
\pkg{brms} (and currently unique to it) is to expand the \code{|}
operator into \code{|<ID>|}, where \code{<ID>} can be any value or
symbol. Person or item parameters with the same \code{ID} will then be
modeled as correlated even though they appear in different \proglang{R}
formulas. That is, if we want to model both person and item parameters
as correlated, respectively, across all distributional parameters, we
choose some arbitray IDs, for instance \texttt{p} for person and
\texttt{i} for item. Importantly, \texttt{p} and \texttt{i} are no
variables in the dataset, they are simply symbols used to connect
multiple varying effects of the same grouping factor (\texttt{person} or
\texttt{item} in this case). For example, we can write

\begin{CodeChunk}

\begin{CodeInput}
R> formula = bf(
R>   y ~ 1 + (1 |i| item) + (1 |p| person),
R>   par2 ~ 1 + (1 |i| item) + (1 |p| person),
R>   par3 ~ 1 + (1 |i| item) + (1 |p| person),
R>   ...
R> )
\end{CodeInput}
\end{CodeChunk}

Because we used \texttt{\textbar{}p\textbar{}} in all person-related
terms, this implies that person parameters (of the same person) are
modeled as correlated across the three distributional parameters
\texttt{mu} (the main parameter implicit in the above formula),
\texttt{par2}, and \texttt{par3}. The same logic applies to item
parameters.

As discussed above, standard \proglang{R} formula syntax is designed to
create additive predictors by splitting up the right-hand side of the
\texttt{formula} in its unique terms separated from each other by
\texttt{+} signs. This formulation is convenient and flexible but it
cannot be used to express non-linear predictors of arbitrary complexity.
To achieve the latter, \pkg{brms} also features a second, more
expressive way to parse \proglang{R} formulas. Suppose that the response
\texttt{y} is related to some covariate \texttt{x} via a non-linear
function \texttt{fun}. Further, suppose that the form of \texttt{fun} is
determined by two parameters \texttt{nlpar1} and \texttt{nlpar2} which
we need to estimate as part of the model fitting process. I will call
them \emph{non-linear parameters} to refer to the fact that they are
parameters of a non-linear function. To complicate things,
\texttt{nlpar1} and \texttt{nlpar2} are not necessarily constant across
observations, but instead may vary across persons and item. That is, we
need to specify a main non-linear formula as well as some additional
linear formulas describing how the non-linear parameters are predicted
by person and item parameters. Basically, non-linear parameters are
handled in the same way as distributional parameters. Suppose that
\texttt{nlpar1} depends on both persons and items, while \texttt{nlpar2}
just depends on the items. In \pkg{brms}, we can express this as

\begin{CodeChunk}

\begin{CodeInput}
R> formula = bf(
R>   y ~ fun(x, nlpar1, nlpar2),
R>   nlpar1 ~ 1 + (1 | item) + (1 | person),
R>   nlpar2 ~ 1 + (1 | item),
R>   nl = TRUE
R> )
\end{CodeInput}
\end{CodeChunk}

Using \texttt{nl\ =\ TRUE} is essential as it ensures that the
right-hand side of the formula is taken literally instead of being
parsed via standard \proglang{R} formula syntax. Of course, we are not
limited to one covariate and two non-linear parameters, but instead are
able to specify any number of them in the formula. Further, the linear
predictors of the non-linear parameters may contain all kinds of
additive terms that I introduced above for usage with distributional
parameters. This combination of linear and non-linear formulas results
in a great amount of model flexibility for the purpose of IRT modeling.

\hypertarget{brms-prior}{%
\subsection{Specifying the prior argument}\label{brms-prior}}

Prior specification is an essential part in the Bayesian workflow and
\pkg{brms} offers an intuitive and flexible interface for convenient
prior specification that can be readily applied to IRT models. In the
following, I explain the syntax to specify priors in the proposed IRT
framework. The priors I choose as examples below are not meant to
represent any specific practical recommendations. Rather, the prior can
only be understood in the context of the model it is a part of
\citep{gelman2017}. Accordingly, user-defined priors should always be
chosen by keeping the model and relevant subject matter knowledge in
mind. I will attempt to provide more ideas in this direction in Section
\ref{examples}.

The main function for the purpose of prior specification in \pkg{brms}
is \texttt{set\_prior}. It takes the prior itself in the form of a
character string as well as additional arguments to define the
parameters on which the prior should be imposed. If we use partial
pooling for item and/or person parameters, the normal prior on those
parameters is automatically set and cannot be changed via the
\texttt{prior} argument. However, we may change priors on the
hyperparameters defining the covariance matrix of the person or item
parameters that is on the standard deviations and correlation matrices.
Suppose we want to define a \(\text{half-Cauchy}(0, 5)\) prior on the
standard devation \(\sigma_\theta\) of the person parameters and an
\(\text{LKJ}(2)\) prior on their correlation matrix
\(\mathbf{\Omega_\theta}\) across the whole model, then we write

\begin{CodeChunk}

\begin{CodeInput}
R> prior = set_prior("cauchy(0, 5)", class = "sd", group = "person") +
R>   set_prior("lkj(2)", class = "cor", group = "person")
\end{CodeInput}
\end{CodeChunk}

These priors will then apply to all distributional and non-linear
parameters which vary across persons. As shown above, multiple priors
may be combined via the \texttt{+} sign. Alternatively, \texttt{c()} or
\texttt{rbind()} may be used to combine priors too. In \proglang{Stan},
and therefore also in \pkg{brms}, truncated priors such as the
half-Cauchy prior are implicitly specified by imposing a hard boundary
on the parameter, that is a lower boundary of zero for standard
deviations, and then using the non-truncated version of the prior.
Setting the hard boundary is done internally and so
\texttt{"cauchy(...)"} will actually imply a half-Cauchy prior when used
for a standard deviation parameter.

We can make priors specific to certain distributional parameters by
means of the \texttt{dpar} argument. For instance, if we want a
\(\text{Gamma}(1, 1)\) prior on the person standard deviation of
\texttt{dpar2} we write

\begin{CodeChunk}

\begin{CodeInput}
R> prior = set_prior("gamma(1, 1)", class = "sd", group = "person", 
R>                   dpar = "dpar2")
\end{CodeInput}
\end{CodeChunk}

Analogously to distributional parameters, priors can be applied
specifically to certain non-linear parameters by means of the
\texttt{nlpar} argument.

Parameters can be fixed to constants by using the \texttt{constant()}
function inside \texttt{set\_prior}. For example, if we want to fix the
standard deviation of person parameters to \(1\) in order to ensure
strong identification of a binary 2PL model, we can write:

\begin{CodeChunk}

\begin{CodeInput}
R> prior = set_prior("constant(1)", class = "sd", group = "person")
\end{CodeInput}
\end{CodeChunk}

If one chooses to \emph{not} use partial pooling for the item parameters
via formulas like
\texttt{y\ \textasciitilde{}\ 0\ +\ item\ +\ (1\ \textbar{}\ person)},
that is, apply non-hierarchical priors, item parameters will be treated
as ordinary regression coefficients and so their prior specification
changes too. In this case, we are not limited to setting priors on all
item parameters, but may also specify them differentially for certain
items if desired. In \pkg{brms}, the class referring to regression
coefficients if called \texttt{"b"}. That is, we can impose a
\(\text{Normal}(0, 3)\) prior on all item parameters via

\begin{CodeChunk}

\begin{CodeInput}
R> prior = set_prior("normal(0, 3)", class = "b")
\end{CodeInput}
\end{CodeChunk}

We may additionally set priors on the specific items. If, say, we know
that \texttt{item1} will be relatively easy to answer correctly, we may
encode this via a prior that has a mean greater than
zero\footnote{Rememember that \pkg{brms} uses
the easiness formulation so that larger values mean higher probability of
solving an item.}. This could then look as follows:

\begin{CodeChunk}

\begin{CodeInput}
R> prior = set_prior("normal(0, 3)", class = "b") +
R>   set_prior("normal(2, 3)", class = "b", coef = "item1")
\end{CodeInput}
\end{CodeChunk}

Internally, \pkg{brms} will always search for the most specific prior
provided by the user. If no user-specified prior can be found, default
priors will apply which are set to be very wide and can thus be
considered non or weakly informative. Priors on the covariates can be
specified in the same way as priors on non-hierarchical item parameters,
that is via class \code{"b"}.

\hypertarget{stan-code-generation}{%
\subsection{Stan code generation}\label{stan-code-generation}}

With a few exceptions outside the scope of this paper, the trinity of
\texttt{formula}, \texttt{family}, and \texttt{prior} completely defines
the model structure and generated \proglang{Stan} code. For example, the
\proglang{Stan} code of the 1PL model with non-hierarchical item priors
generated from

\begin{CodeChunk}

\begin{CodeInput}
R> make_stancode(
R>   formula = y ~ 0 + item + (1 | person),
R>   family = brmsfamily("bernoulli", "logit"),
R>   prior = prior(normal(0, 5), class = "b") +
R>     prior(normal(0, 3), class = "sd", group = "id"),
R>   ...
R> )
\end{CodeInput}
\end{CodeChunk}

looks as follows:

\begin{CodeChunk}

\begin{CodeOutput}
// generated with brms 2.11.5
functions {
}
data {
  int<lower=1> N;  // number of observations
  int Y[N];  // response variable
  int<lower=1> K;  // number of population-level effects
  matrix[N, K] X;  // population-level design matrix
  // data for group-level effects of ID 1
  int<lower=1> N_1;  // number of grouping levels
  int<lower=1> M_1;  // number of coefficients per level
  int<lower=1> J_1[N];  // grouping indicator per observation
  // group-level predictor values
  vector[N] Z_1_1;
  int prior_only;  // should the likelihood be ignored?
}
transformed data {
}
parameters {
  vector[K] b;  // population-level effects
  vector<lower=0>[M_1] sd_1;  // group-level standard deviations
  vector[N_1] z_1[M_1];  // standardized group-level effects
}
transformed parameters {
  vector[N_1] r_1_1;  // actual group-level effects
  r_1_1 = (sd_1[1] * (z_1[1]));
}
model {
  // initialize linear predictor term
  vector[N] mu = X * b;
  for (n in 1:N) {
    // add more terms to the linear predictor
    mu[n] += r_1_1[J_1[n]] * Z_1_1[n];
  }
  // priors including all constants
  target += normal_lpdf(b | 0, 5);
  target += normal_lpdf(sd_1 | 0, 3)
    - 1 * normal_lccdf(0 | 0, 3);
  target += normal_lpdf(z_1[1] | 0, 1);
  // likelihood including all constants
  if (!prior_only) {
    target += bernoulli_logit_lpmf(Y | mu);
  }
}
generated quantities {
}
\end{CodeOutput}
\end{CodeChunk}

As can be seen, the \proglang{Stan} code is heavily commented to
facilitate human readability. There are a lot of steps involved in order
to fully generate the \proglang{Stan} code and I will explain the most
important steps below.

In a first step, the information in \texttt{family} and \texttt{formula}
is combined so that all \proglang{R} formulas can be validated against
the model family and its supported distributional parameters. The
formula of each distributional parameter is then split into several
predictor terms, for our purposes most notably population-level
(`overall') and group-level (`varying') terms. In the 1PL example, this
means splitting the right-hand side of
\texttt{y\ \textasciitilde{}\ 0\ +\ item\ +\ (1\ \textbar{}\ person)}
into \texttt{\textasciitilde{}\ 0\ +\ item} (population-level terms) and
\texttt{\textasciitilde{}\ (1\ \textbar{}\ person)} (group-level terms).
Their primary difference between the two is that non-hierarchical priors
are applied to the former while hierarchical priors are applied to the
latter. The generated \proglang{Stan} code varies substantially between
these two types of terms in particular to exploit the sparsity of the
group-level parameters (each observation belongs only to a single level
per grouping factor). In order to model group-level parameters as
correlated across different model parts, each being represented by its
own formula, the corresponding group-level terms are extracted from the
formulas and combined into a single object which can then be handled
independently of the individual formulas. The \proglang{Stan} code
generation then proceeds as follows.

Data and parameters of group-level terms are added to the \texttt{data},
\texttt{parameters}, \texttt{transformed\ parameters}, and
\texttt{model} block \citep[see][ for more details on \proglang{Stan}
programming blocks]{carpenter2017}. The most important aspect of the
generated \proglang{Stan} code is the use of the non-centered
parameterization for group-level coefficients. That is, instead of
defining the coefficients \texttt{r\_*} as \texttt{parameters} and
directly applying hierarchical priors, we define independent standard
normal coefficients \texttt{z\_*} as \texttt{parameters} and then scale
them according to the hyperparameters (i.e., standard deviations
\texttt{sd\_*} and Cholesky factors of correlation matrices
\texttt{L\_*} if required) in \texttt{transformed\ parameters} . For a
single group-level term, that is, without having to take correlations
into account, this happens via

\begin{verbatim}
r_1_1 = (sd_1[1] * (z_1[1]));
\end{verbatim}

In the \proglang{Stan} code, matrices are split into their column
vectors whenever indexing the individual columns over observations is
required. This leads to more efficient sampling in most cases. The two
suffix numbers of the variable names indicate the grouping factor number
and term number within a grouping factor, respectively. These suffixes
are replaced by more informative names after model fitting. For example,
\texttt{r\_1\_1} refers to the coefficient vector of the first
group-level term within with first grouping factor.

For each individual model component/formula, the \texttt{data},
\texttt{parameters} and \texttt{model} blocks of the \proglang{Stan}
code are written independently. A vector named after the predicted
distributional parameter of length equal to the number of observations
is created in the \texttt{model} block and filled with predictions.
Predictor terms are added to this vector in two steps. First, everything
that can be efficiently expressed via (non-sparse) matrix algebra, is
added to the predictor for all observations at once. For example,
population-level terms, such as non-hierarchical item parameters or
covariate effects, are computed by matrix-multiplying their design
matrix \texttt{X} with the related coefficients \texttt{b}. For the main
parameter \texttt{mu}, the following line of \proglang{Stan} code is
generated:

\begin{verbatim}
mu = X * b;
\end{verbatim}

All remaining terms are added in a loop over observations. For a single
group-level term (e.g., hierarchical person parameters) the related
\proglang{Stan} code is

\begin{verbatim}
for (n in 1:N) {
  mu[n] += r_1_1[J_1[n]] * Z_1_1[n];
}
\end{verbatim}

In the above code, \texttt{Z\_1\_1} is the data vector of the variable
whose effect is supposed to vary (equal to \texttt{1} if only a
group-level intercept is present), \texttt{r\_1\_1} is the vector of
group-level coefficients, and \texttt{J\_1} is the index vector linking
group-levels to observations.

The \texttt{family} argument is used to define the overall form of the
likelihood. Together with \texttt{formula}, the most efficient
\proglang{Stan} expression for the likelihood is found. Specific
combinations of likelihood distribution and link function yield simplied
and more efficient mathematical forms. \proglang{Stan} has several of
these built-in combinations, for example, the
\texttt{bernoulli\_logit\_lpmf} function, where \texttt{lpmf} stands for
`log probability density function'. If such an optimized function is
available, it is used in the generated \proglang{Stan} code. Otherwise,
the regular \texttt{lpmf} function for the distribution is used (e.g.,
\texttt{bernoulli\_lpmf}) and the link function is applied manually
beforehand. Next, it is determined whether the likelihood can be
vectorized over observations. This drastically speeds up estimation time
if the likelihood contains terms which are constant across observations
and thus need to be computed only once instead of separately for each
observation. As \proglang{Stan} directly compiles to \proglang{C++},
loops themselves are fast and so the reason for vectorizing in
\proglang{Stan} is different from the reason to vectorize in
\proglang{R} or another interpreted programming language. For the 1PL
model above, the likelihood is written in \proglang{Stan} a follows:

\begin{verbatim}
target += bernoulli_logit_lpmf(Y | mu);
\end{verbatim}

The \texttt{target\ +=} notation is used to indicate that a term is
added to the log-posterior. If, for some reason, we could not have
vectorized over the likelihood, the code would have looked as follows,
instead:

\begin{verbatim}
for (n in 1:N) {
  target += bernoulli_logit_lpmf(Y[n] | mu[n]);
}
\end{verbatim}

Lastly, the priors are added to the \proglang{Stan} code also using the
\texttt{target\ +=} syntax. The \texttt{normal(0,\ 5)} prior on the item
parameters is translated to \proglang{Stan} as follows:

\begin{verbatim}
target += normal_lpdf(b | 0, 5);
\end{verbatim}

The (half) \texttt{normal(0,\ 3)} prior on the person parameters'
standard deviation is specified as

\begin{verbatim}
target += normal_lpdf(sd_1 | 0, 3) - 1 * normal_lccdf(0 | 0, 3);
\end{verbatim}

The term \texttt{-\ 1\ *\ normal\_lccdf(0\ \textbar{}\ 0,\ 3)} is
constant across parameters and, hence, does not affect parameter
estimation. The only reason for its presence in Stan models written by
\pkg{brms} is that it enables correct marginal likelihood estimation
after model fitting via the \texttt{bridgesampling} package
\citep{bridgesampling}. It does so by correcting the log-posterior for
the truncation of the prior, which was automatically and implicitly done
by \proglang{Stan} since \texttt{sd\_1} was defined with a lower
boundary of \texttt{0} in the \texttt{parameters} block. If marginal
likelihood estimation, is not of interest, adding such constants is not
necessary and can be safely ignored.

\hypertarget{estimation}{%
\section{Parameter estimation and post-processing}\label{estimation}}

The \pkg{brms} package uses \proglang{Stan} \citep{carpenter2017} on the
back-end for the model estimation. Accordingly, all samplers implemented
in \proglang{Stan} can be used to fit \pkg{brms} models. The flagship
algorithm of \proglang{Stan} is an adaptive Hamiltonian Monte-Carlo
(HMC) sampler \citep{betancourt2014, betancourt2017, stanM2019}, which
represents a progression from the No-U-Turn Sampler (NUTS) by
\citet{hoffman2014}. HMC-like algorithms produce posterior samples that
are much less autocorrelated than those of other samplers such as the
random-walk Metropolis algorithm \citep{hoffman2014, creutz1988}. What
is more, consecutive samples may even be anti-correlated leading to
higher efficiency than completely independent samples
\citep{vehtari2019}. The main drawback of this increased efficiency is
the need to calculate the gradient of the log-posterior, which can be
automated using algorithmic differentiation \citep{griewank2008}, but is
still a time-consuming process for more complex models. Thus, using HMC
leads to higher quality samples but takes more time per sample than
other typically applied algorithms. Another drawback of HMC is the need
to pre-specify at least two parameters, which are both critical for the
performance of HMC. The adaptive HMC Sampler of \proglang{Stan} allows
setting these parameters automatically thus eliminating the need for any
hand-tuning, while still being at least as efficient as a well tuned HMC
\citep{hoffman2014}. For more details on the sampling algorithms applied
in \proglang{Stan}, see the \proglang{Stan} user's manual
\citep{stanM2019} as well as \citet{hoffman2014}.

After the estimation of the parameters' joint posterior distribution,
\pkg{brms} offers a wide range of post-processing options of which
several are helpful in an IRT context. Below, I introduce the most
important post-processing options. I will show their usage in hands-on
examples in the upcoming sections. For a quick numerical and graphical
summary, respectively, of the central model parameters, I recommend the
\code{summary} and \code{plot} methods. The posterior distribution of
person parameters (and, if also modeled as varying effects, item
parameters) can be extracted with the \code{coef} method. The
\code{hypothesis} method can be used to easily compute and evaluate
parameter contrasts, for instance, when the goal is to compare the
difficulty of two items or the ability of two persons. A visualization
of the effects of item or person covariates is readily available via the
\code{conditional_effects} method.

With the help of the \code{posterior_predict} method, \pkg{brms} allows
drawing samples from the posterior predictive distribution. This not
only allows to make predictions for existing or new data, but also
enables the comparison between the actual response \(y\) and the
response \(\hat{y}\) predicted by the model. Such comparisons can be
visualized in the form of posterior-predictive checks by means of the
\code{pp_check} method \citep{gabry2019}. Further, via the
\code{log_lik} method, the pointwise log-likelihood can be obtained,
which can be used, among others, for various cross-validation methods.
One widely applied cross-validation approach is leave-one-out
cross-validation \citep[LOO-CV;][]{vehtari2017loo}, for which an
approximate version is available via the \code{loo} method of the
\pkg{loo} package \citep{vehtari2017loo, vehtari2017psis}. If LOO-CV is
not an option or if the approximation fails, exact k-fold
cross-valdiation is available via the \code{kfold} method. The
cross-validation results can be further post-processed for the purpose
of comparison, selection, or averaging of models. In these contexts, the
\code{loo_compare}, \code{model_weights}, and \code{pp_average} methods
are particularily helpful.

In addition to cross-validation based fit measures, the marginal
likelihood (i.e., the denomintor in Bayes' theorem) and marginal
likelihood ratios, commonly known as Bayes factors, can be used for
model comparison, selection, or averaging as well \citep{kass1995}. In
general, obtaining the marginal likelihood of a model is a
computationally demanding task \citep{kass1995}. In \pkg{brms}, this is
realized via bridgesampling \citep{meng1996, meng2002} as implemented in
the \pkg{bridgesampling} package \citep{bridgesampling}. The
corresponding methods are called \code{bridge_sampler} to obtain (log)
marginal likelihood estimates, \code{bayes_factor} to obtain Bayes
factors and \code{post_prob} to obtain posterior model probabilities
based on prior model probabilities and marginal likelihood estimates.

\hypertarget{examples}{%
\section{Examples}\label{examples}}

In this section, I am going to discuss several examples of advanced IRT
models that can be fitted with \pkg{brms}. I will focus on three common
model classes: binary, ordinal, and reaction time models, but the
discussed principles also apply to other types of responses that can be
analyzed by means of IRT.

\hypertarget{binary}{%
\subsection{Binary Models}\label{binary}}

To illustrate the application of \pkg{brms} to binary IRT models, I will
use the \code{VerbAgg} data set \citep{deboeck2004}, which is included
in the \pkg{lme4} package \citep{lme4}.

\begin{CodeChunk}

\begin{CodeInput}
R> data("VerbAgg", package = "lme4")
\end{CodeInput}
\end{CodeChunk}

This data set contains responses of 316 participants on 24 items of a
questionnaire on verbal aggression. Several item and person covariates
are provided. A glimpse of the data is given in Table
\ref{tab:head-VerbAgg} and more details can be found by typing
\code{?lme4::VerbAgg}.

\begin{CodeChunk}
\begin{table}

\caption{\label{tab:head-VerbAgg}First ten rows of the \texttt{VerbAgg} data.}
\centering
\begin{tabular}[t]{rllllllll}
\toprule
Anger & Gender & item & resp & id & btype & situ & mode & r2\\
\midrule
20 & M & S1WantCurse & no & 1 & curse & other & want & N\\
11 & M & S1WantCurse & no & 2 & curse & other & want & N\\
17 & F & S1WantCurse & perhaps & 3 & curse & other & want & Y\\
21 & F & S1WantCurse & perhaps & 4 & curse & other & want & Y\\
17 & F & S1WantCurse & perhaps & 5 & curse & other & want & Y\\
\addlinespace
21 & F & S1WantCurse & yes & 6 & curse & other & want & Y\\
39 & F & S1WantCurse & yes & 7 & curse & other & want & Y\\
21 & F & S1WantCurse & no & 8 & curse & other & want & N\\
24 & F & S1WantCurse & no & 9 & curse & other & want & N\\
16 & F & S1WantCurse & yes & 10 & curse & other & want & Y\\
\bottomrule
\end{tabular}
\end{table}

\end{CodeChunk}

Let us start by computing a simple 1PL model. For reasons discussed in
Section \ref{model}, I prefer to partially pool person and item
parameters by specifying the model as

\begin{CodeChunk}

\begin{CodeInput}
R> formula_va_1pl <- bf(r2 ~ 1 + (1 | item) + (1 | id))
\end{CodeInput}
\end{CodeChunk}

If we wanted to specify non-hierarchical item parameters, instead, we
would have had to use the formula

\begin{CodeChunk}

\begin{CodeInput}
R> bf(r2 ~ 0 + item + (1 | id))
\end{CodeInput}
\end{CodeChunk}

To impose a small amount of regularization on the model, I set
\(\text{half-normal}(0, 3)\) priors on the hierarchical standard
deviations of person and items parameters. Given the scale of the
logistic response function, this can be regarded as a weakly informative
prior.

\begin{CodeChunk}

\begin{CodeInput}
R> prior_va_1pl <- 
R+   prior("normal(0, 3)", class = "sd", group = "id") + 
R+   prior("normal(0, 3)", class = "sd", group = "item")
\end{CodeInput}
\end{CodeChunk}

The model is then fit as follows:

\begin{CodeChunk}

\begin{CodeInput}
R> fit_va_1pl <- brm(
R>   formula = formula_va_1pl,
R>   data = VerbAgg, 
R>   family = brmsfamily("bernoulli", "logit"),
R>   prior = prior_va_1pl
R> )
\end{CodeInput}
\end{CodeChunk}

To get a quick overview of the model results and convergence, we can
summarize the main parameters numerically using the \texttt{summary}
method:

\begin{CodeChunk}

\begin{CodeInput}
R> summary(fit_va_1pl)
\end{CodeInput}

\begin{CodeOutput}
 Family: bernoulli 
  Links: mu = logit 
Formula: r2 ~ 1 + (1 | item) + (1 | id) 
   Data: VerbAgg (Number of observations: 7584) 
Samples: 4 chains, each with iter = 2000; warmup = 1000; thin = 1;
         total post-warmup samples = 4000

Group-Level Effects: 
~id (Number of levels: 316) 
              Estimate Est.Error l-95% CI u-95% CI Rhat Bulk_ESS Tail_ESS
sd(Intercept)     1.39      0.07     1.25     1.54 1.00      906     1441

~item (Number of levels: 24) 
              Estimate Est.Error l-95% CI u-95% CI Rhat Bulk_ESS Tail_ESS
sd(Intercept)     1.20      0.19     0.89     1.62 1.00      531     1326

Population-Level Effects: 
          Estimate Est.Error l-95% CI u-95% CI Rhat Bulk_ESS Tail_ESS
Intercept    -0.16      0.26    -0.71     0.34 1.03      196      532

Samples were drawn using sampling(NUTS). For each parameter, Bulk_ESS
and Tail_ESS are effective sample size measures, and Rhat is the potential
scale reduction factor on split chains (at convergence, Rhat = 1).
\end{CodeOutput}
\end{CodeChunk}

A graphical summary of the marginal posterior densities as well as the
MCMC chains is obtained via

\begin{CodeChunk}

\begin{CodeInput}
R> plot(fit_va_1pl)
\end{CodeInput}
\begin{figure}

{\centering \includegraphics{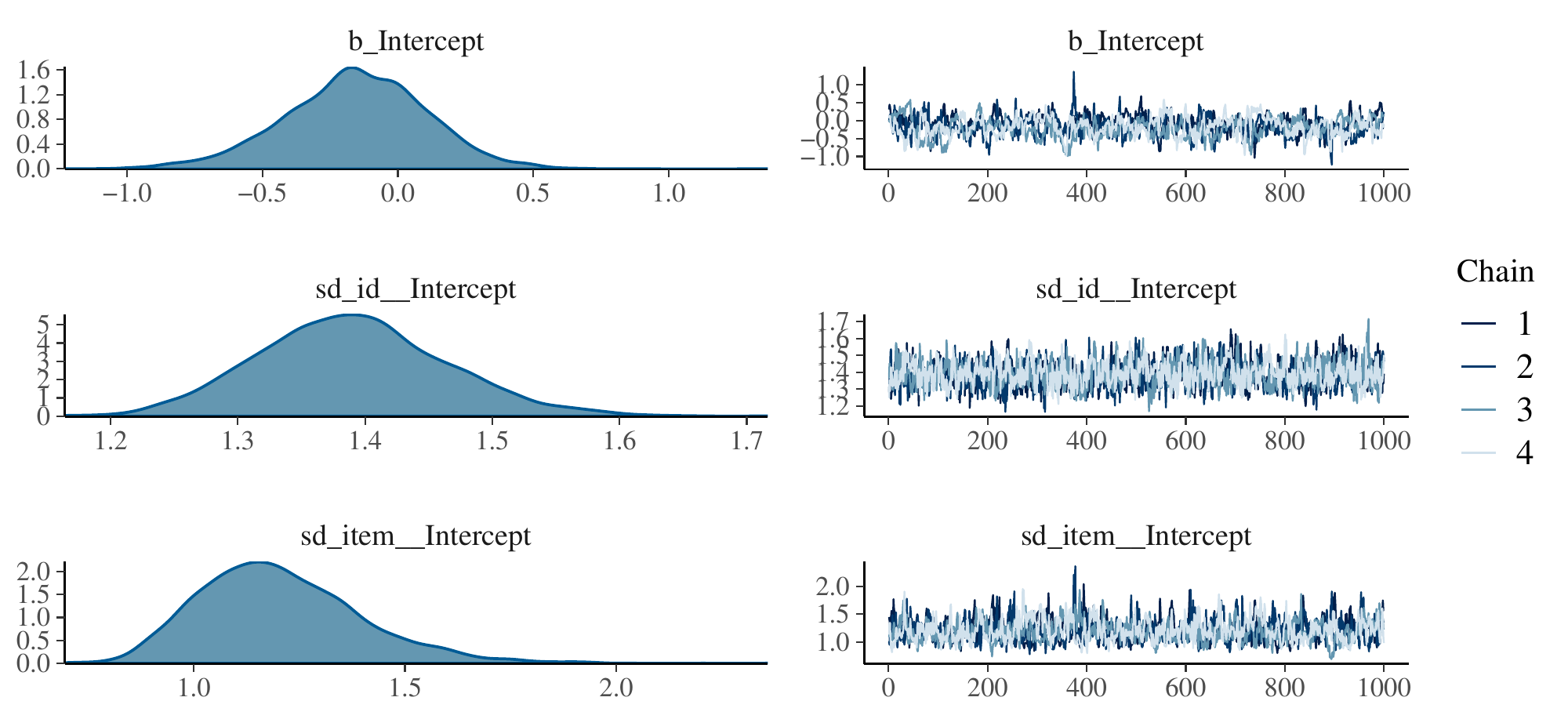} 

}

\caption{Summary of the posterior distribution of selected parameters obtained by model \code{fit\_va\_1pl.}}\label{fig:plot-va-1pl}
\end{figure}
\end{CodeChunk}

and shown in Figure \ref{fig:plot-va-1pl}. Before interpreting the
results, it is crucial to investigate whether the model fitting
algortihm converged to its target, that is, the parameters' posterior
distribution for fully Bayesian models. There are multiple ways to
investigate convergence. We could do so graphically by looking at trace
plots (see the right-hand side of Figure \ref{fig:plot-va-1pl}) or more
recently proposed rank plots \citep{vehtari2019}. On that basis, we can
interpret MCMC chains as having converged to the same target
distribution, if the chains are mixing well individually (i.e., quickly
jumping up and down) and are overlaying one another at the same time
\citep{gelman2013}. We may also investigate convergence numerically by
means of the scale reduction factor \(\widehat{R}\)
\citep{gelman1992, gelman2013, vehtari2019}, which should be close to
one (i.e., \(\widehat{R} < 1.05\)), and the effective sample size, which
should be as large as possible but at least 400 to merely ensure
reliable convergence diagnostics \citep{vehtari2019}. The corresponding
columns in the summary output are called \code{Rhat} and
\code{Eff.Sample}. Convergence diagnostics for all model parameters can
be obtained via the \code{rhat} and \code{neff_ratio} methods,
respectively. Additionally, there are some diagnostics specific to
(adaptive) HMC, which we can access using \code{nuts_params} and plotted
via various options in \code{mcmc_plot}. After investigating both the
graphical and numerical indicators of convergence, we are confident that
the model fitting algorithm succeeded so that we can start interpreting
the results.

We see from the summary of the standard deviation parameters (named
\texttt{sd(intercept)} in the output) that both persons and items vary
substantially. Not all model parameters are shown in \texttt{summary}
and \texttt{plot} to keep the output clean and readable and so we need
to call other methods depending on what we are interested in. In IRT,
this most likely includes the person and item parameters, which we can
access via methods \texttt{coef} and \texttt{ranef} depending on whether
or not we want to include overall effects (i.e., the global intercept
for the present model) in the computation of the individual
coefficients. This would typically be the case if we were interested in
obtaining estimates of item difficulty or person ability. Item and
person parameters are displayed in Figure \ref{fig:coef-item-va-1pl} and
\ref{fig:coef-person-va-1pl}, respectively.

\begin{CodeChunk}
\begin{figure}

{\centering \includegraphics{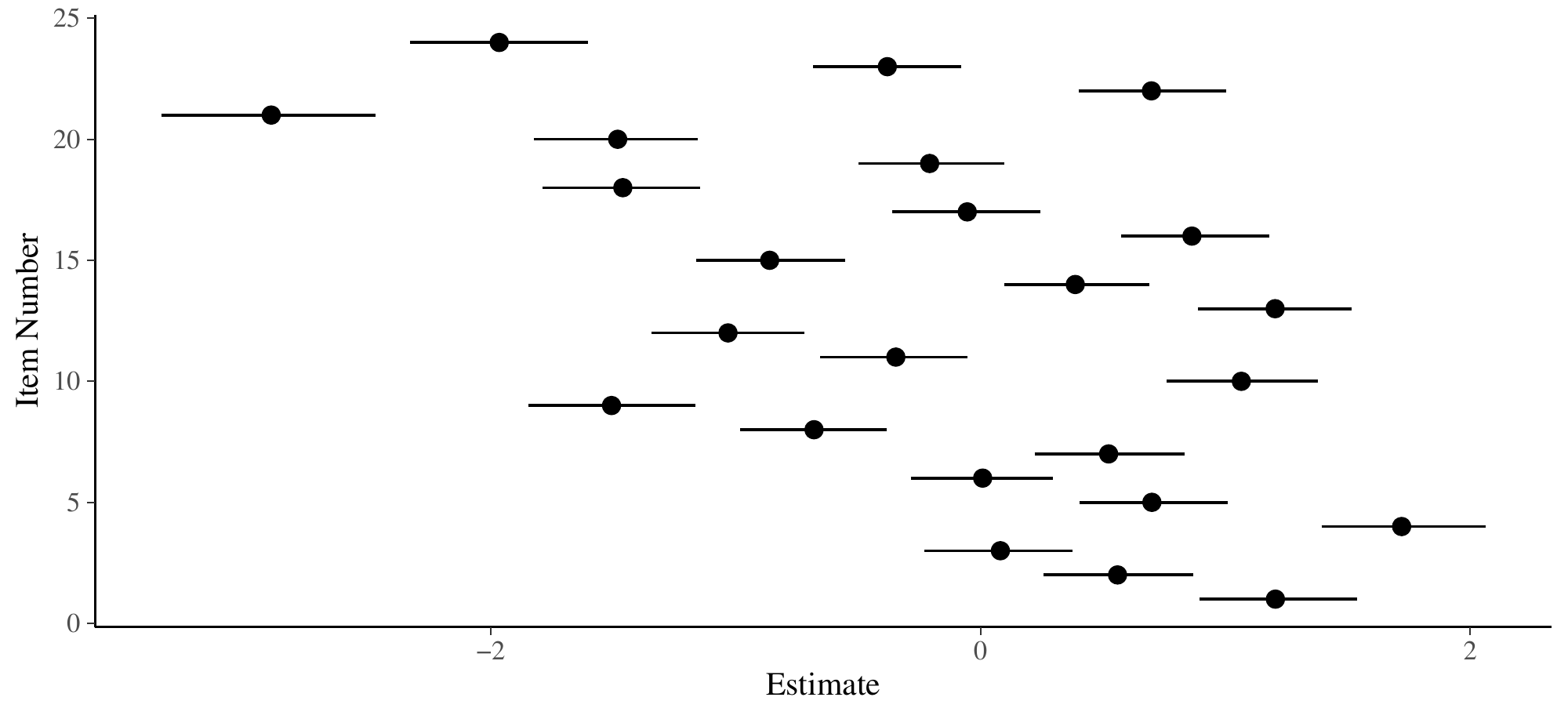} 

}

\caption[Posterior means and 95\% credible intervals of item parameters as estimated by model \code{fit\_va\_1pl}]{Posterior means and 95\% credible intervals of item parameters as estimated by model \code{fit\_va\_1pl}.}\label{fig:coef-item-va-1pl}
\end{figure}
\end{CodeChunk}

\begin{CodeChunk}
\begin{figure}

{\centering \includegraphics{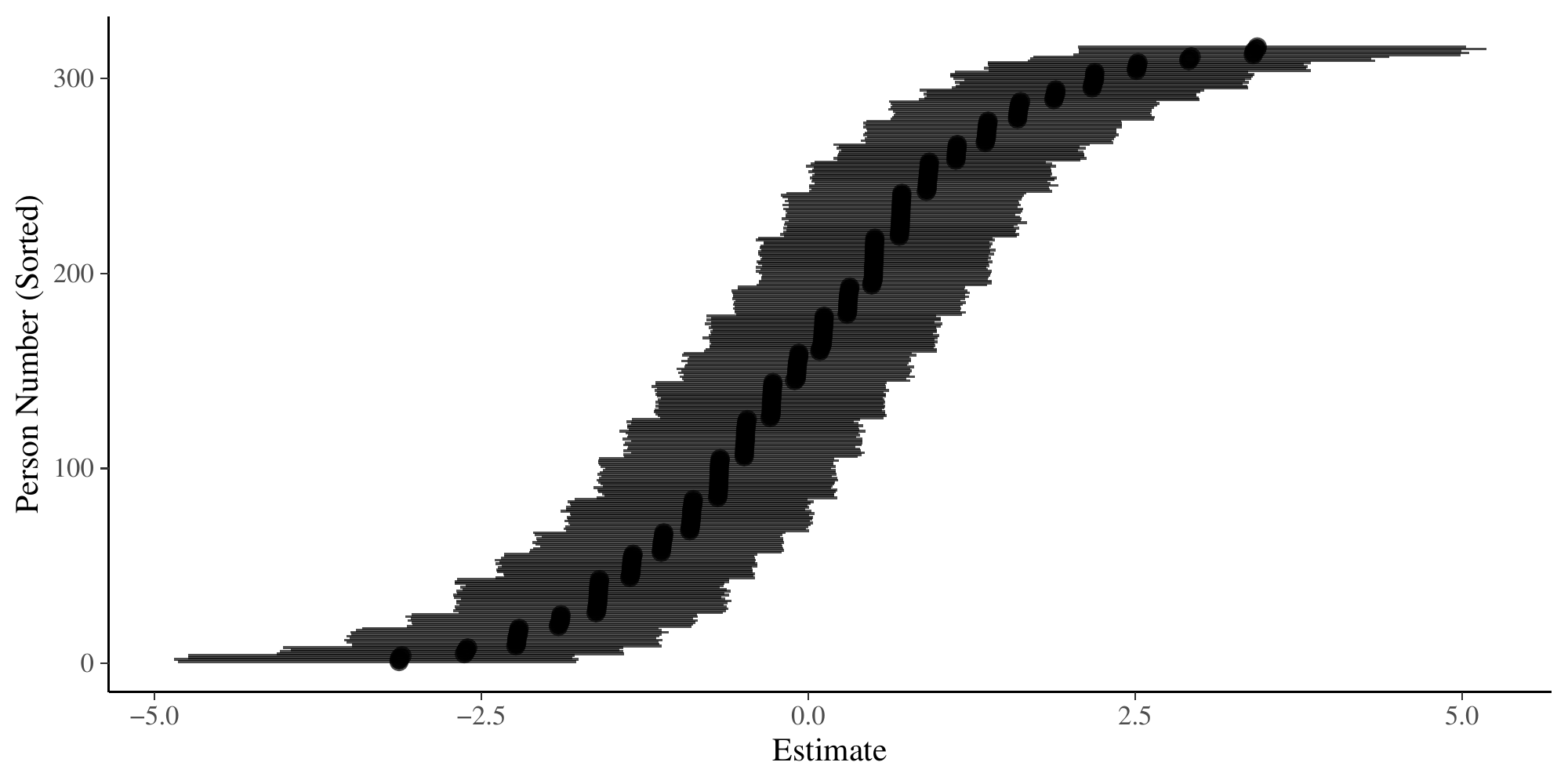} 

}

\caption[Posterior means and 95\% credible intervals of person parameters (sorted) as estimated by model \code{fit\_va\_1pl}]{Posterior means and 95\% credible intervals of person parameters (sorted) as estimated by model \code{fit\_va\_1pl}.}\label{fig:coef-person-va-1pl}
\end{figure}
\end{CodeChunk}

From Figure \ref{fig:coef-item-va-1pl} it is clear that some items
(e.g., the 4th item) are agreed on by a lot of individuals and thus have
strongly positive easieness parameters, while other items (e.g., the
21th item) are mostly rejected and thus have a strongly negative
easiness parameter. From Figure \ref{fig:coef-person-va-1pl} we see that
the person parameters vary a lot but otherwise show a regular pattern of
blocks of persons getting very similar estimates (given the same prior).
The latter is because, in the 1PL model, all items are assumed to have
the same discrimination and are thus weighted equally. As a result, two
persons endorsing the same number of items in total will receive the
same estimate, regardless of which items they endorsed exactly. This
assumption of equal discriminations is very restrictive and I will now
investigate it in more detail. In a 2PL model, we would assume each item
to have its own discrimination, which are to be estimated from the model
along with all other parameters. Recall that mathematically, the 2PL
model looks as follows:

\[
P(y = 1) = \mu = \text{logistic}(\alpha_i (\theta_p + \xi_i))
\]

Without any further restrictions, this model will likely not be
identified (unless we were specifiying highly informative priors)
because a switch in the sign of \(\alpha_i\) can be corrected for by a
switch in the sign of \(\theta_p + \xi_i\) without a change in the
overall likelihood. For this reason, I assume \(\alpha_i\) to be
positive for all items, a sensible assumption for the \code{VerbAgg}
data set where a \(y = 1\) always implies endorsing a certain verbally
aggressive behavior. There are multiple ways to force \(\alpha_i\) to be
positive, one of which is to model it on the log-scale, that is to
estimate \(\log \alpha_i\) and then exponentiating the result to obtain
the actual discrimination via \(\alpha_i = \exp(\log \alpha_i)\).

\begin{CodeChunk}

\begin{CodeInput}
R> formula_va_2pl <- bf(
R+   r2 ~ exp(logalpha) * eta,
R+   eta ~ 1 + (1 |i| item) + (1 | id),
R+   logalpha ~ 1 + (1 |i| item),
R+   nl = TRUE
R+ )
\end{CodeInput}
\end{CodeChunk}

Above, I split up the non-linear model into two parts, \texttt{eta} and
\texttt{logalpha}, each of which is in turn predicted by a linear
formula. The parameter \texttt{eta} represents the sum of person
parameter and item easiness, whereas \texttt{logalpha} represents the
log discrimination. I modeled item easiness and discrimination as
correlated by using \texttt{\textbar{}i\textbar{}} in both varying item
terms (see Section \ref{brms}). I impose weakly informative priors both
on the intercepts of \texttt{eta} and \texttt{logalpha} (i.e., on the
overall easiness and log discrimination) as well as on the standard
deviations of person and item parameters.

\begin{CodeChunk}

\begin{CodeInput}
R> prior_va_2pl <- 
R+   prior("normal(0, 5)", class = "b", nlpar = "eta") +
R+   prior("normal(0, 1)", class = "b", nlpar = "logalpha") +
R+   prior("constant(1)", class = "sd", group = "id", nlpar = "eta") + 
R+   prior("normal(0, 3)", class = "sd", group = "item", nlpar = "eta") +
R+   prior("normal(0, 1)", class = "sd", group = "item", nlpar = "logalpha")
\end{CodeInput}
\end{CodeChunk}

Finally, I put everything together and fit the model via

\begin{CodeChunk}

\begin{CodeInput}
R> fit_va_2pl <- brm(
R>   formula = formula_va_2pl,
R>   data = VerbAgg, 
R>   family = brmsfamily("bernoulli", "logit"),
R>   prior = prior_va_2pl,
R> ) 
\end{CodeInput}
\end{CodeChunk}

The results of \texttt{summary} and \texttt{plot} indicate good
convergence of the model and I don't show their outputs for brevity's
sake. Instead, I directly take a look at the item parameters in Figure
\ref{fig:coef-item-va-2pl}. The discrimination estimates displayed on
the right-hand have some considerable uncertainty, roughly between 1 and
2, but are overall quite similar across items with posterior mean
estimates of roughly between 1.2 and 1.5. The easiness parameters
displayed on the left-hand side still show a similar pattern as in the
1PL although their estimates a little less spread out as a result of the
discrimination estimates being greater 1.

\begin{CodeChunk}
\begin{figure}

{\centering \includegraphics{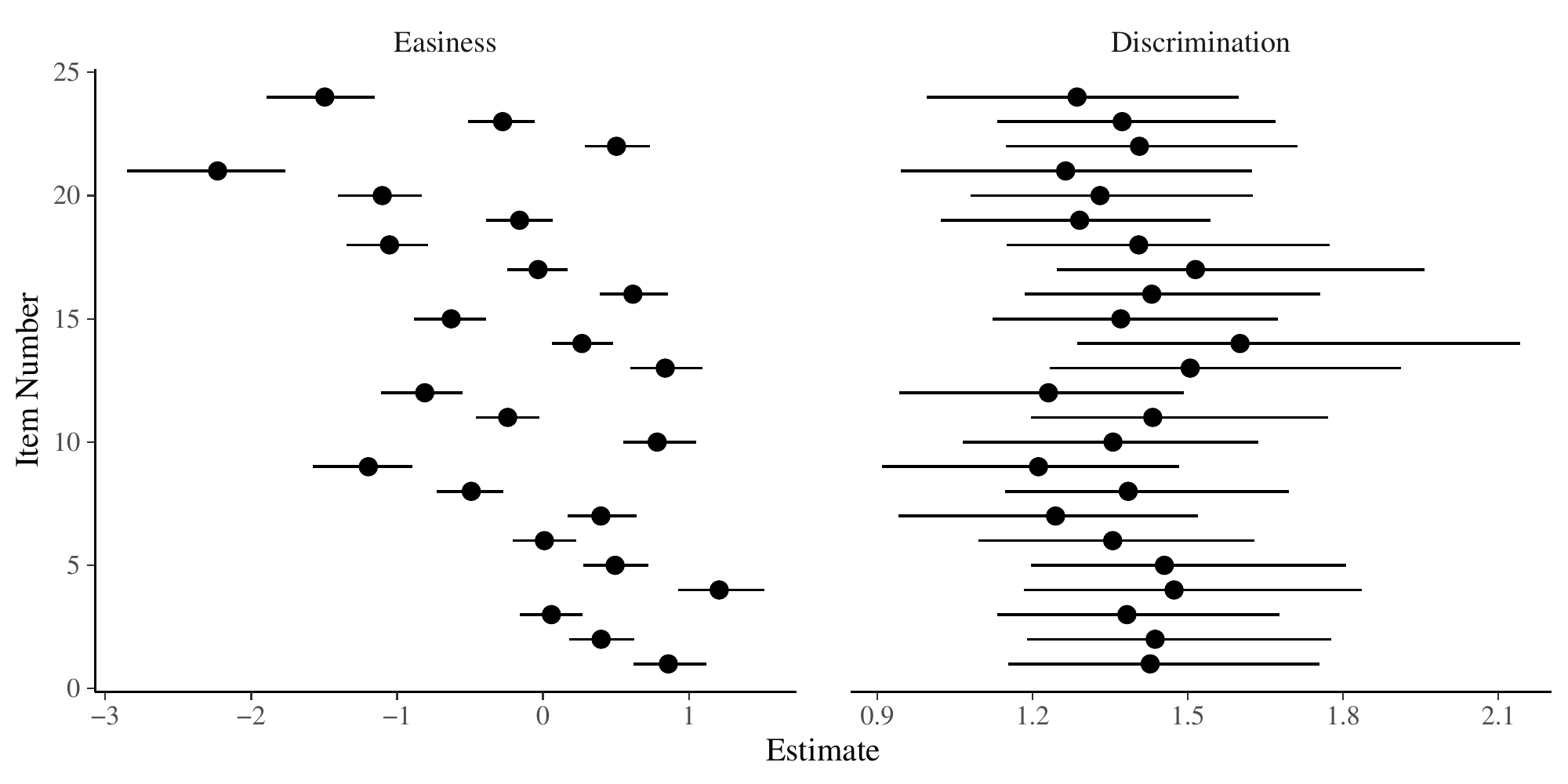} 

}

\caption[Posterior means and 95\% credible intervals of item parameters as estimated by model \code{fit\_va\_2pl}]{Posterior means and 95\% credible intervals of item parameters as estimated by model \code{fit\_va\_2pl}.}\label{fig:coef-item-va-2pl}
\end{figure}
\end{CodeChunk}

\begin{CodeChunk}
\begin{figure}

{\centering \includegraphics{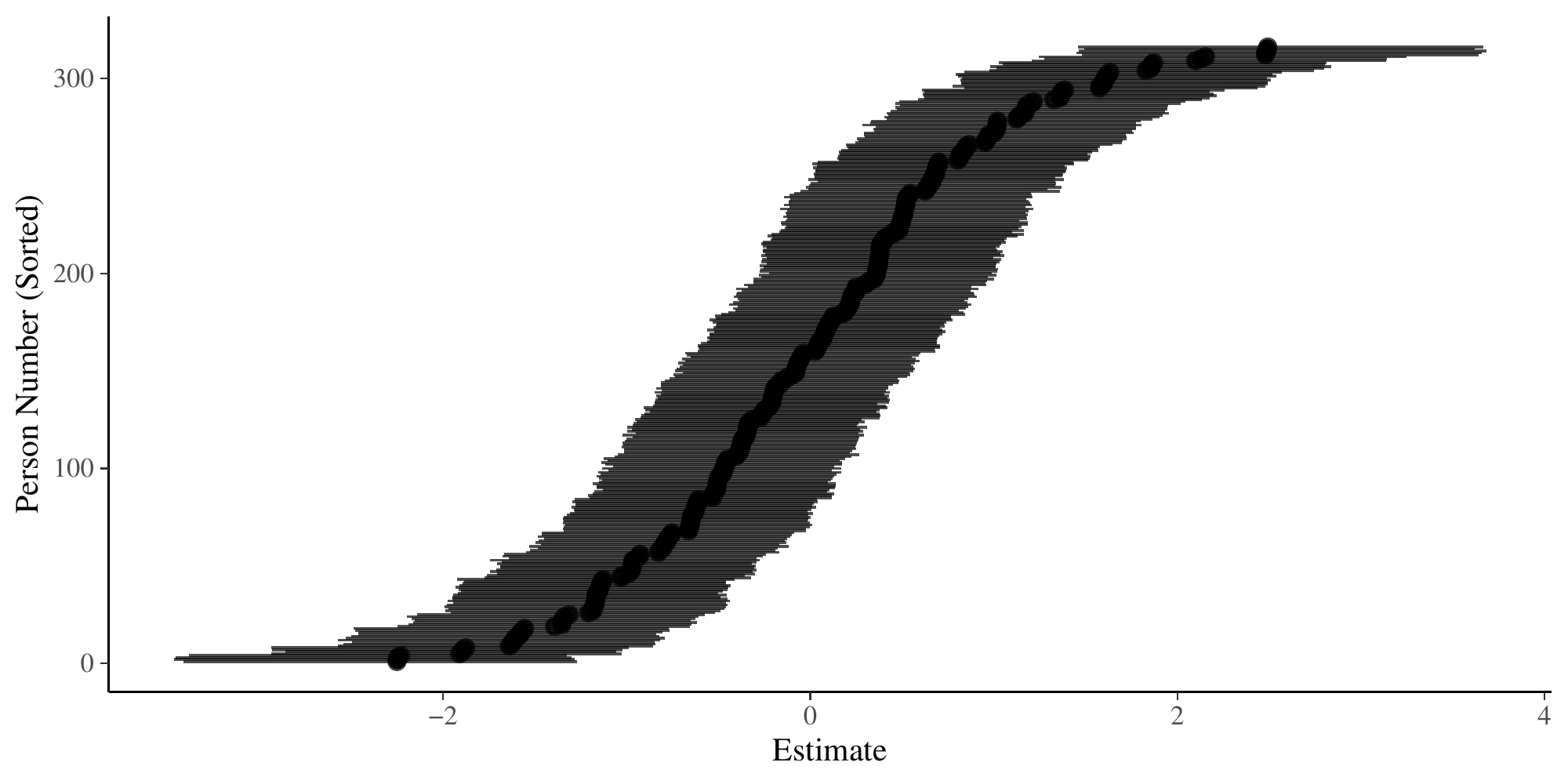} 

}

\caption[Posterior means and 95\% credible intervals of person parameters (sorted) as estimated by model \code{fit\_va\_2pl}]{Posterior means and 95\% credible intervals of person parameters (sorted) as estimated by model \code{fit\_va\_2pl}.}\label{fig:coef-person-va-2pl}
\end{figure}
\end{CodeChunk}

The correlation between person parameters obtained by the two models
turns out to be \(r =\) 1, so there is basically nothing gained from the
2PL model applied to this particular data set. In line with these
results, model fit obtained via approximate leave-one-out
cross-validation (LOO-CV) results in a LOOIC difference of
\(\Delta \text{LOOIC} =\) 5.97 in favor of the 2PL model, which is quite
small both on an absolute scale and in comparison to its standard error
\(\text{SE} =\) 4.78 depicting the uncertainty in the difference. Thus,
I will continue using the 1PL model in the following.

\hypertarget{modeling-covariates}{%
\subsubsection{Modeling Covariates}\label{modeling-covariates}}

When analysing the \code{VerbAgg} data set, I am not so much interested
in the item and person parameters themselves, rather than in the effects
of item and person covariates. I start by including only item
covariates, in this case the behavior type (\texttt{btype}, with factor
levels \texttt{curse}, \texttt{scold}, and \texttt{shout}), the
situation type (\texttt{stype}, with factor levels \texttt{other} and
\texttt{self}), as well as the behavior mode (\texttt{mode}, with factor
levels \texttt{want} and \texttt{do}). Additionally, I assume the effect
of \texttt{mode} to vary over persons, that is assume each person to
have their own effect of \texttt{mode}. We specify this model in formula
syntax as

\begin{CodeChunk}

\begin{CodeInput}
R> r2 ~ btype + situ + mode + (1 | item) + (1 + mode | id)
\end{CodeInput}
\end{CodeChunk}

This model assumes a varying intercept (i.e., baseline) and a varying
effect of \texttt{mode} (i.e., difference between \texttt{want} and
\texttt{do}) per person. However, in this example, I am actually more
interested in estimating varying effects of \texttt{want} and
\texttt{do}, separately, in order to compare variation between these two
modes. For this purpose, we slightly amend the formula, which now
becomes

\begin{CodeChunk}

\begin{CodeInput}
R> r2 ~ btype + situ + mode + (1 | item) + (0 + mode | id)
\end{CodeInput}
\end{CodeChunk}

The notation \texttt{0\ +\ mode} implies that each factor level of
\texttt{mode} gets its own varying effect, instead of modeling the
intercept and differences between factor levels. We are now ready to
actually fit the model:

\begin{CodeChunk}

\begin{CodeInput}
R> formula_va_1pl_cov1 <- bf(
R>   r2 ~ btype + situ + mode + (1 | item) + (0 + mode | id)
R> )
R> fit_va_1pl_cov1 <- brm(
R>   formula = formula_va_1pl_cov1,
R>   data = VerbAgg, 
R>   family = brmsfamily("bernoulli", "logit"),
R>   prior = prior_va_1pl
R> )
\end{CodeInput}
\end{CodeChunk}

As usual, a quick overview of the results can be obtained via

\begin{CodeChunk}

\begin{CodeInput}
R> summary(fit_va_1pl_cov1)
\end{CodeInput}

\begin{CodeOutput}
 Family: bernoulli 
  Links: mu = logit 
Formula: r2 ~ btype + situ + mode + (1 | item) + (0 + mode | id) 
   Data: VerbAgg (Number of observations: 7584) 
Samples: 4 chains, each with iter = 2000; warmup = 1000; thin = 1;
         total post-warmup samples = 4000

Group-Level Effects: 
~id (Number of levels: 316) 
                     Estimate Est.Error l-95% CI u-95% CI Rhat Bulk_ESS Tail_ESS
sd(modewant)             1.47      0.09     1.30     1.64 1.00     1827     2664
sd(modedo)               1.67      0.10     1.48     1.87 1.00     1766     3007
cor(modewant,modedo)     0.77      0.04     0.69     0.84 1.00     1678     2694

~item (Number of levels: 24) 
              Estimate Est.Error l-95% CI u-95% CI Rhat Bulk_ESS Tail_ESS
sd(Intercept)     0.46      0.09     0.31     0.67 1.00     1641     1852

Population-Level Effects: 
           Estimate Est.Error l-95% CI u-95% CI Rhat Bulk_ESS Tail_ESS
Intercept      1.89      0.25     1.42     2.40 1.00     1844     2533
btypescold    -1.13      0.24    -1.62    -0.64 1.00     1938     1973
btypeshout    -2.24      0.25    -2.74    -1.74 1.00     2095     2452
situself      -1.12      0.21    -1.53    -0.71 1.00     2106     2335
modedo        -0.78      0.21    -1.20    -0.38 1.00     2197     2498

Samples were drawn using sampling(NUTS). For each parameter, Bulk_ESS
and Tail_ESS are effective sample size measures, and Rhat is the potential
scale reduction factor on split chains (at convergence, Rhat = 1).
\end{CodeOutput}
\end{CodeChunk}

From the summary output, we see that the behavior difference of the
\texttt{do} and \texttt{want} behavior modes has a negative logit
regression coefficient (\(b =\) -0.78, \(95\% \; \text{CI} = [\)-1.2,
-0.38\(]\)), which implies that, holding other predictors constant,
people are more likely to \emph{want} to be verbally agressive than to
actually \emph{be} verbally agressive. However, although the direction
of the effect is quite clear, its magnitude tends to be hard to
interprete as it the regression coefficients are on the logit scale. To
ease interpretation, we can transform and plot them on the original
probability scale (see Figure \ref{fig:me-va-mode}) using a single line
of code:

\begin{CodeChunk}

\begin{CodeInput}
R> conditional_effects(fit_va_1pl_cov1, "mode")
\end{CodeInput}
\begin{figure}

{\centering \includegraphics{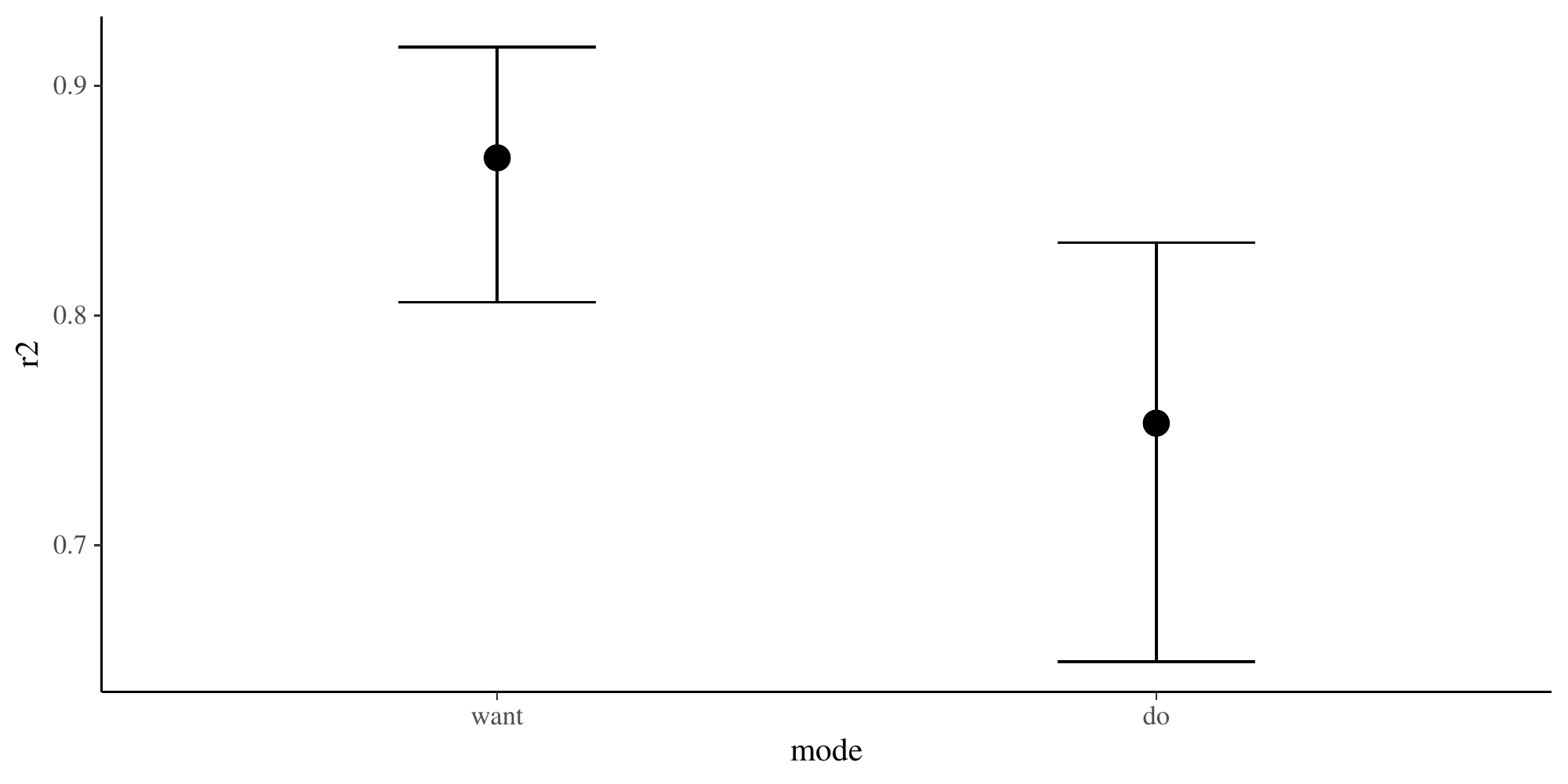} 

}

\caption[Expected probabilities of agreeing to an item in the VerbAgg data set as a function of the behavior mode conditioned on all other covariates being set to their reference categories]{Expected probabilities of agreeing to an item in the VerbAgg data set as a function of the behavior mode conditioned on all other covariates being set to their reference categories.}\label{fig:me-va-mode}
\end{figure}
\end{CodeChunk}

Further, in the summary output, we see that both modes vary
substantially over persons, with a little bit more variation in mode
\texttt{do}. We may ask the question how likely it is, that the
variation in \texttt{do} across persons is actually larger than the
variation in \texttt{want}. Answering such a question in a frequentist
framework would not be easy as the joint distribution of the two SD
parameters is unlikely to be (bivariate) normal. In contrast, having
obtained samples from the joint posterior distribution using MCMC
sampling, as we did, computing the posterior distribution of the
difference becomes a matter of computing the difference for each pair of
posterior samples. This procedure of transforming posterior samples is
automated in the \texttt{hypothesis} method of \pkg{brms}. For this
particular question, we need to use it as follows:

\begin{CodeChunk}

\begin{CodeInput}
R> hyp <- "modedo - modewant > 0"
R> hypothesis(fit_va_1pl_cov1, hyp, class = "sd", group = "id")
\end{CodeInput}
\end{CodeChunk}

\begin{CodeChunk}

\begin{CodeOutput}
             Hypothesis Estimate CI.Lower CI.Upper Post.Prob
1 (modedo-modewant) > 0      0.2     0.02     0.39      0.97
\end{CodeOutput}
\end{CodeChunk}

(output shortend for readability; CI denotes \(90\%\) the credibly
interval). From the \texttt{Post.Prob} column we see that, given the
model and the data, with 0.97 probability the SD of the \texttt{do}
effects is higher than the SD of the \texttt{want} effects, although the
expected SD difference of 0.2 (on the logit scale) is rather small.

Similarily to how we incorporate item covariates, we may also add person
covariates to the model. In the \code{VerbAgg} data, we have information
about the subjects' trait anger score \texttt{Anger} as measured on the
Stat Trait Anger Expression Inventory \citep[STAXI;][]{spielberger2010}
as well as about their \texttt{Gender}. Let us additionally assume
\texttt{Gender} and \texttt{mode} to interact, that is allowing the
effect of the behavior mode (\texttt{do} vs.~\texttt{want}) to vary with
the gender of the subjects. Further, I expect the individual item
parameters to also vary with gender by replacing the term
\texttt{(1\ \textbar{}\ item)} by
\texttt{(0\ +\ Gender\ \textbar{}\ item)}. The complete model formula
then looks as follows:

\begin{CodeChunk}

\begin{CodeInput}
R> r2 ~ Anger + Gender + btype + situ + mode + mode:Gender +
R>   (0 + Gender | item) + (0 + mode | id)
\end{CodeInput}
\end{CodeChunk}

We fit the model as usual with the \texttt{brm} function. Afterwards, we
obtain a graphical summary of the effects of the newly added person
covariates via

\begin{CodeChunk}

\begin{CodeInput}
R> conditional_effects(fit_va_1pl_cov2, c("Anger", "mode:Gender"))
\end{CodeInput}
\end{CodeChunk}

As visible on the left-hand side of Figure \ref{fig:me-va-cov2},
increased trait anger is clearly associated with higher probabilities of
agreeing to items in the \code{VerbAgg} data set. Also, as can be seen
on the right-hand side of Figure \ref{fig:me-va-cov2}, there is an
interaction between behavior mode and gender. More specifically, women
and men report wanting to be verbally aggressive by roughly the same
probability, while men report actually being verbally aggressive with a
much higher probability than women.

\begin{CodeChunk}
\begin{figure}

{\centering \includegraphics{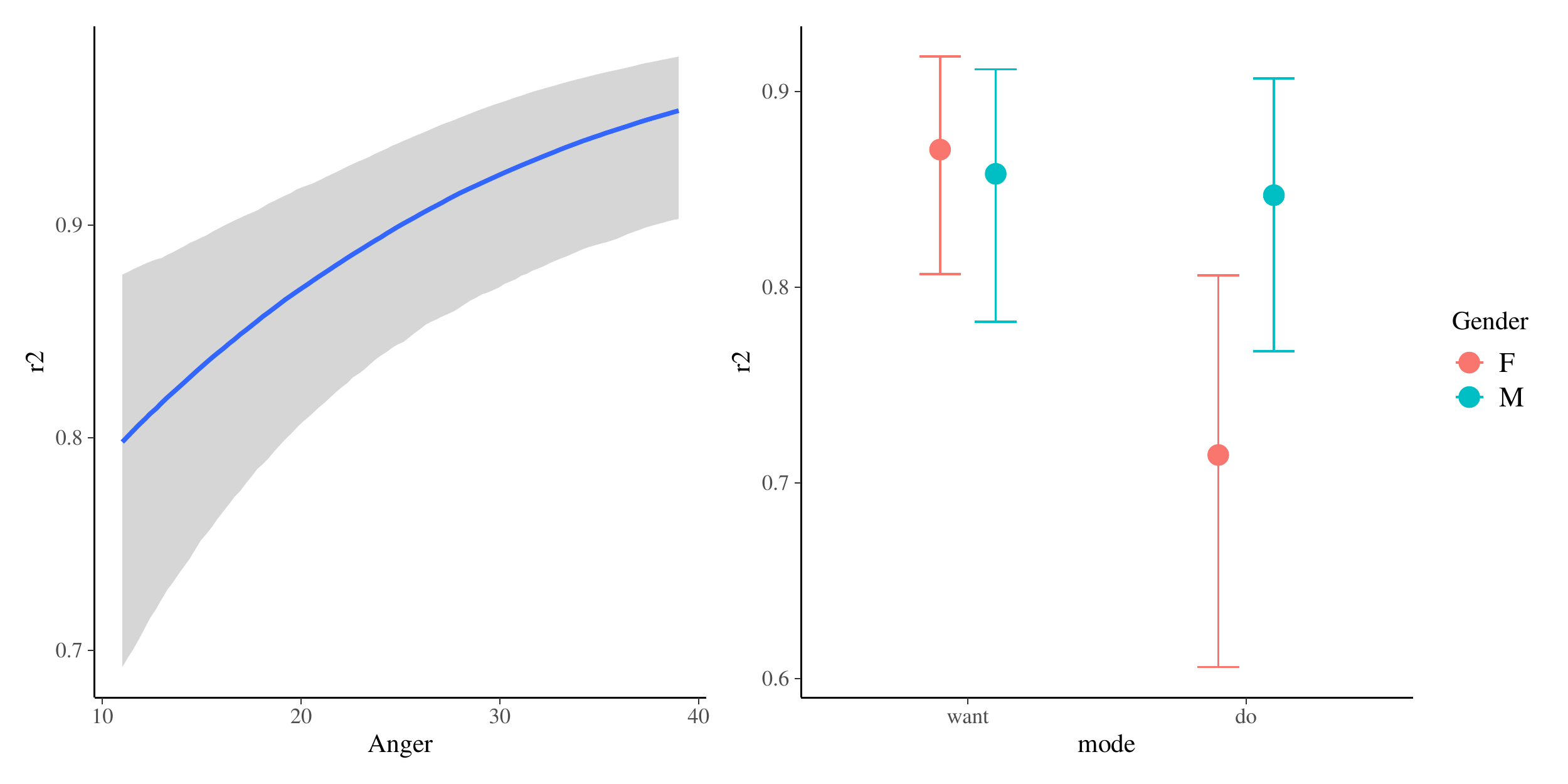} 

}

\caption[Expected probabilities of agreeing to an item in the VerbAgg data set as a function of the trait anger (left) and the interaction of behavior mode and subjects' gender (right) conditioned on all other categorical covariates being set to their reference categories and numerical covariates being set to their mean]{Expected probabilities of agreeing to an item in the VerbAgg data set as a function of the trait anger (left) and the interaction of behavior mode and subjects' gender (right) conditioned on all other categorical covariates being set to their reference categories and numerical covariates being set to their mean.}\label{fig:me-va-cov2}
\end{figure}
\end{CodeChunk}

The modeled interaction between behavior mode and gender can be also
understood as a DIF analysis to investigate whether the item property
`behavior mode' has differential implications depending on the person's
gender. Indeed, the above results indicate the existence of such DIF.
Let us run some more fine grained DIF analysis and investigate whether
items involving actual cursing or scolding (but not shouting) show DIF
with respect to gender \citep[inspired by][]{deboeck2011}. For this
purpose, we create a dummy \texttt{dif} variable that is \texttt{1} for
females on items involving actual cursing or scolding and \texttt{0}
otherwise:

\begin{CodeChunk}

\begin{CodeInput}
R> VerbAgg$dif <- as.numeric(with(
R+   VerbAgg, Gender == "F" & mode == "do" & btype %in% c("curse", "scold")
R+ ))
\end{CodeInput}
\end{CodeChunk}

This new person-by-item covariate is then used as a predictor in
addition to \texttt{Gender}. Including the latter is important so that
the \texttt{dif} variable does not capture variation due to mean
differences in gender, which are not an indication of DIF. The formula
then looks as follows:

\begin{CodeChunk}

\begin{CodeInput}
R> r2 ~ Gender + dif + (1 | item) + (1 | id)
\end{CodeInput}
\end{CodeChunk}

The regression coefficient of \texttt{dif} is estimated as \(b =\) -0.94
(95\% CI = {[}-1.22, -0.66{]}), which indicates that woman (say they)
curse and scold more rarely than men. As noted by \citet{deboeck2011},
this analysis tests DIF of the uniform type, that is DIF which is
independent of the specific value of the latent trait.

In all of the covariate models described above, there is no particular
reasoning behind the choice of which item or person covariates are
assumed to vary over persons or items, respectively, and which are
assumed to be constant. We may also try to model multiple or even all
item covariates as varying over persons and all person covariates as
varying over items. In fact, this maximal multilevel approach may be
more robust and conservative \citep{barr2013}. In frequentist
implementations of multilevel models, we often see convergence issues
when using maximal multilevel structure \citep{bates2015}. This has been
interpreted by some as an indication of overfitting \citep{bates2015}
while others disagree \citep{barr2013}. In any case, convergence issues
seem to be a crude indicator of overfitting that I argue should not be
blindly relied on. Fortunately, convergence of complex multilevel models
turns out to be much less of a problem when using gradient-based MCMC
samplers such as HMC \citep{hoffman2014}. For instance, when fitting a
maximal multilevel structure of item and person covariates via the
formula

\begin{CodeChunk}

\begin{CodeInput}
R> r2 ~ 1 + Anger + Gender + btype + situ + mode + 
R>   (1 + Anger + Gender | item) + (1 + btype + situ + mode  | id)
\end{CodeInput}
\end{CodeChunk}

the \pkg{lme4} package indicates serious convergence issues while the
\pkg{brms} model converges just fine (results not displayed here, see
the supplementary \proglang{R} code for details). Of course, this is not
to say that such a multilevel structure is necessarily sensible.
However, being able to fit those models allows for more principled ways
of testing afterwards \emph{if} the assumed complexity is actually
supported by the data, for instance via cross-validation or Bayes
factors.

\hypertarget{modeling-guessing-parameters}{%
\subsubsection{Modeling Guessing
Parameters}\label{modeling-guessing-parameters}}

A common aspect of binary item response data in IRT is that persons may
be able to simply guess the correct answer with a certain non-zero
probability. This may happen in a forced choice format where the correct
answer is presented along with some distractors. As a result, the
probability of correctly answering an item never falls below the
guessing probability, regardless of the person's ability. For instance,
when assuming all alternatives to be equally attractive in the absense
of any knowledge about the correct answer, the guessing probability is 1
divided by the total number of alternatives. Such a property of the
administered items needs to be taken into account in the estimated IRT
model. The most commonly applied model in such a situation is the 3PL
model\footnote{In addition to guessing probabilities, which increase the lower
bound of success probability beyond 0, it is also possible that lapses decrease
the upper bound of the sucess probability below 1. A binary model taking into
account both guesses and lapses is referred to 4PL model. Arguably 4PL models
more relevant for instance in psychophycis and less so in IRT. For that reason,
I do not discuss it in more detail in this paper but want to point out that
\pkg{brms} could also be used to fit 4PL models.}. Mathematically, the
model can be expressed as

\[
P(y = 1) = \mu = \gamma_i + (1 - \gamma_i) \times \text{logistic}(\alpha_i (\theta_p + \xi_i))
\]

where \(\gamma_i\) represents the guessing probability of item \(i\) and
all other parameters have the same meaning as in the 2PL model.

The items of the \code{VerbAgg} data set do not have a forced choice
response format -- and no right or wrong answers either -- and so
modeling guessing probabilities makes little sense for that data. For
brevity's sake, I am not going to introduce another data set on which I
apply 3PL models, but instead only focus on showing how to express such
a model in \pkg{brms} without actually fitting the model. For an
application of 3PL models to real data using \pkg{brms}, see
\citet{buerkner2020spm}.

Suppose we have administered forced choice items with 4 response
alternatives of which only one is correct, then -- under the assumption
of equal probabilities of choosing one of the alternatives in case of
guessing -- we obtain a guessing probability of \(25\%\). When modeling
this guessing probability as known and otherwise following the
recommendation presented in Section \ref{model}, we can write down the
formula of the 3PL model as follows:

\begin{CodeChunk}

\begin{CodeInput}
R> formula_va_3pl <- bf(
R+   r2 ~ 0.25 + 0.75 * inv_logit(exp(logalpha) * eta),
R+   eta ~ 1 + (1 |i| item) + (1 | id),
R+   logalpha ~ 1 + (1 |i| item),
R+   nl = TRUE
R+ )
R> family_va_3pl <- brmsfamily("bernoulli", link = "identity")
\end{CodeInput}
\end{CodeChunk}

Above, I incorporated the logistic response function directly into the
formula via \texttt{inv\_logit}. As a result, the predictions of the
overall success probabilities are already on the right scale and thus an
additional usage of a link function is neither required nor reasonable.
In other words, we have to apply the \texttt{identity} link function. Of
course, we may also add covariates to all linear predictor terms of the
model (i.e., to \texttt{eta} and \texttt{logalpha}) in the same way as
demonstrated above for the 2PL model.

If we did not know the guessing probabilities, we can decide to estimate
them along with all other model paramters. In \pkg{brms} syntax, the
model then looks as follows:

\begin{CodeChunk}

\begin{CodeInput}
R> formula_va_3pl <- bf(
R+   r2 ~ gamma + (1 - gamma) * inv_logit(exp(logalpha) * eta),
R+   eta ~ 1 + (1 |i| item) + (1 | id),
R+   logalpha ~ 1 + (1 |i| item),
R+   logitgamma ~ 1 + (1 |i| item),
R+   nlf(gamma ~ inv_logit(logitgamma)),
R+   nl = TRUE
R+ )
\end{CodeInput}
\end{CodeChunk}

There are some important aspects of this model specification that
require further explanation. Since \(\gamma_i\) is a probability
parameter, we need to restrict it between \(0\) and \(1\). One solution
is to model \(\gamma_i\) on the logit scale via
\texttt{logitgamma\ \textasciitilde{}\ 1\ +\ (1\ \textbar{}i\textbar{}\ item)}
and then transform it back to the original scale via the
\texttt{inv\_logit} function, which exists both in \pkg{brms} and in
\proglang{Stan}. I could have done this directly in the main formula but
this would have implied doing the transformation twice, as
\texttt{gamma} appears twice in the formula. For increased efficiency, I
have defined both \texttt{gamma} and \texttt{logitgamma} as non-linear
parameters and related them via\\
\texttt{gamma\ \textasciitilde{}\ inv\_logit(logitgamma)}. Passing the
formula to \texttt{nlf} makes sure that the formula for \texttt{gamma}
is treated as non-linear in the same way as setting \texttt{nl\ =\ TRUE}
does for the main formula.

There are some general statistical problems with the 3PL model with
estimated guessing probabilities, because the interpretability of the
model parameters, in particular of the item difficulty and
discrimination, suffers as a result \citep{han2012}. Accordingly, it may
be more favorable to design items with known guessing probabilities in
the first place.

\hypertarget{ordinal}{%
\subsection{Ordinal Models}\label{ordinal}}

When analysing the \code{VerbAgg} data using binary IRT models, I have
assumed participants responses on the items to be a dichotomous
\texttt{yes} vs.~\texttt{no} decision. However, this is actually not
entirely accurate as the actual responses were obtained on an ordinal
three-point scale with the options \texttt{yes}, \texttt{perhaps},
\texttt{no}. In the former section, I have combined \texttt{yes} and
\texttt{perhaps} into one response category, following the analysis
strategy of \citet{deboeck2011}. In \pkg{brms}, we are not bounded to
reducing the response to a binary decision but are instead able to use
the full information in the response values by applying ordinal IRT
models. There are multiple ordinal model classes
\citep{agresti2010, buerkner2019}, one of which is the graded response
model (GRM; see Section \ref{respdists}). As a reminder, when modeling
the responses \(y\) via the GRM, we do not only have a predictor term
\(\eta\), but also a vector \(\tau\) of \(C-1\) ordered latent
thresholds, where \(C\) is the number of categories (\(C = 3\) for the
\code{VerbAgg} data). The GRM assumes that the observed ordinal
responses arise from the categorization of a latent continuous variable,
which is predicted by \(\eta\). The thresholds \(\tau\) indicate those
latent values where the observable ordinal responses change from one to
another category. An illustration of the model's assumptions is provied
in Figure \ref{fig:grm}.

\begin{CodeChunk}
\begin{figure}

{\centering \includegraphics{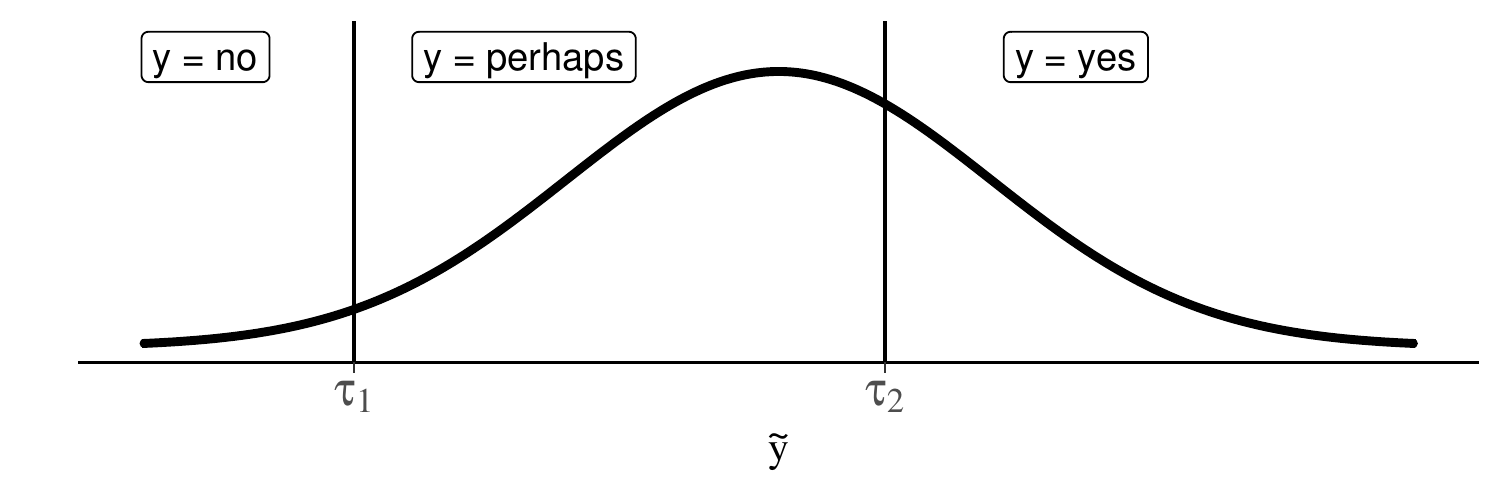} 

}

\caption{Assumptions of the graded response model when applied to the \code{VerbAgg} data. The area under the curve in each bin represents the probability of the corresponding event given the set of possible events for the latent variable $\tilde{y}$, which depends linearily on the predictor term $\eta$.}\label{fig:grm}
\end{figure}
\end{CodeChunk}

The model specification of the GRM, or for that matter of any ordinal
model class, is quite similar to binary models. For now, the only
changes are that we switch out the binary variable \texttt{r2} in favor
of the three-point ordinal variable \texttt{resp} and use the
\texttt{cumulative} instead of the \texttt{bernoulli} family:

\begin{CodeChunk}

\begin{CodeInput}
R> formula_va_ord_1pl <- bf(resp ~ 1 + (1 | item) + (1 | id))
R> fit_va_ord_1pl <- brm(
R>   formula = formula_va_ord_1pl,
R>   data = VerbAgg,
R>   family = brmsfamily("cumulative", "logit"),
R>   prior = prior_va_1pl
R> )
\end{CodeInput}
\end{CodeChunk}

With regard to the ordinal thresholds, this implies modeling an overall
threshold vector shared across items plus an item-specific constant (via
\texttt{(1\ \textbar{}\ item)}) that is added to the overall threshold
vector. In other words, for this model, the thresholds of two items are
simply shifted to the left or right relative to each other but otherwise
share the same shape. This is a quite restrictive assumption and I will
discuss a generalization of it later on.

The \texttt{summary} and \texttt{plot} output look very similar to the
ones from the binary model except for that we now see two intercepts,
which represent the ordinal thresholds. I do not show their outputs here
for brevity's sake. Instead, let us focus on what exactly has changed in
the estimation of the person parameters. As displayed on the left-hand
side of Figure \ref{fig:va-person-scatter}, person parameters estimated
by the binary and those estimated by the ordinal model are largely in
alignment with each other although we can observe bigger differences for
larger values. The latter is to be expected since, in the ordinal model,
I kept the two higher categories \texttt{perhaps} and \texttt{yes}
separate thus increasing the information for larger but not so much for
smaller person parameters. In accordance with this observation, we see
that the person parameters whose precision has increased the most
through the usage of an ordinal model are those with large mean values
(see right-hand side of Figure \ref{fig:va-person-scatter}). Taken
together, we clearly gain something from correctly treating the response
as ordinal, not only theoretically -- \texttt{perhaps} is certainly
something else than \texttt{yes} in most people's mind -- but also
statistically by increasing the precision of the estimates.

\begin{CodeChunk}
\begin{figure}

{\centering \includegraphics{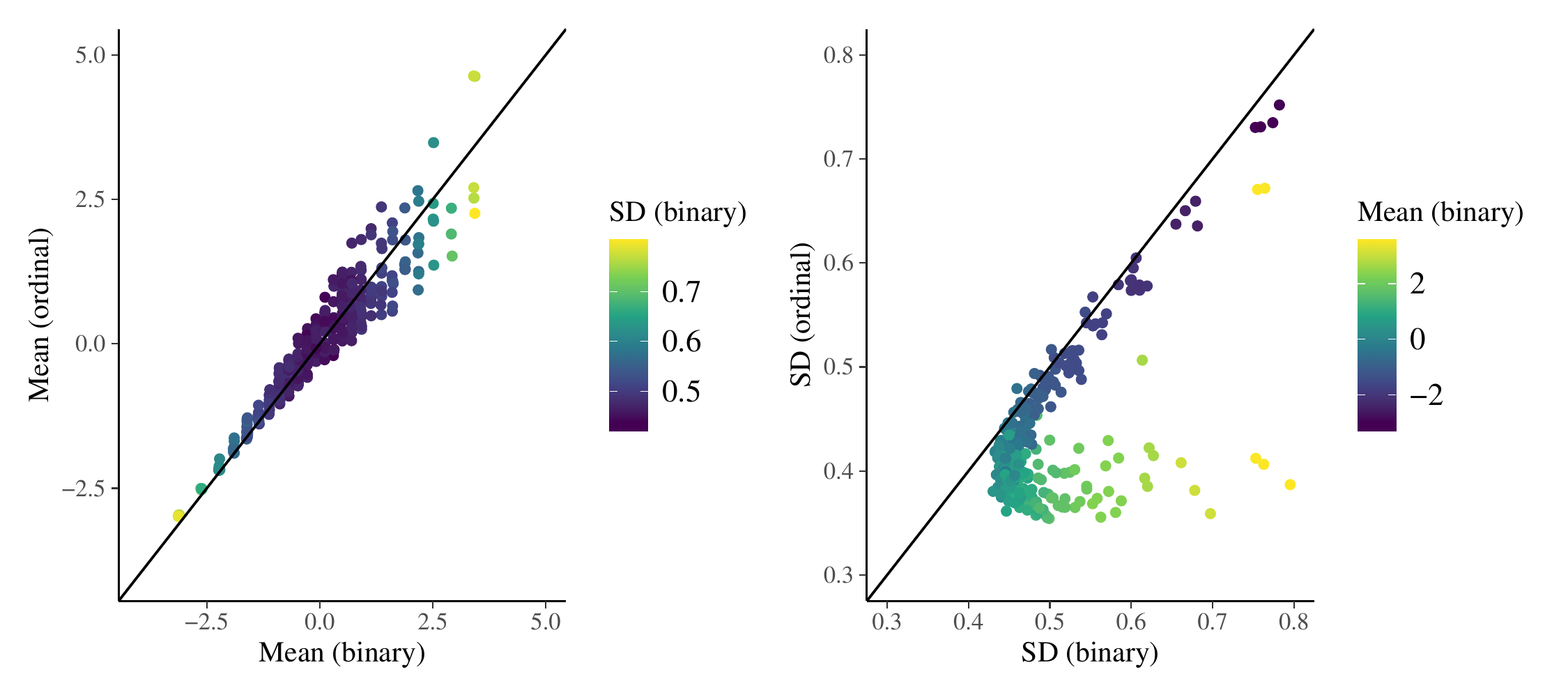} 

}

\caption[Relationship of person parameters estimated by the binary 1PL model and the ordinal graded response model]{Relationship of person parameters estimated by the binary 1PL model and the ordinal graded response model. Posterior means are shown on the left-hand side and Posterior standard deviations are shown on the right-hand side.}\label{fig:va-person-scatter}
\end{figure}
\end{CodeChunk}

As mentioned above, the assumption of shifted but equally shaped
threshold vectors across items is quite restrictive. To test this
assumption, we fit a second ordinal model in which each item receives
its own threshold vector so that threshold locations and shapes are
completely free to vary across items. Doing so in the
\texttt{cumulative} family currently requires to use non-hiearchical
priors on the thresholds as otherwise their order requirements may not
be satisfied \citep{buerkner2019}. Recently, an order preserving
generalization to hierarchical priors has been proposed and may
eventually find its way into brms \citep{paulewicz2020}. The model with
fully varying thresholds across items can be specified as follows:

\begin{CodeChunk}

\begin{CodeInput}
R> formula_va_ord_thres_1pl <- bf(resp | thres(gr = item) ~ 1 + (1 | id))
R> prior_va_ord_thres_1pl <- 
R+  prior("normal(0, 3)", class = "Intercept") + 
R+  prior("normal(0, 3)", class = "sd", group = "id")
\end{CodeInput}
\end{CodeChunk}

\begin{CodeChunk}

\begin{CodeInput}
R> fit_va_ord_thres_1pl <- brm(
R>  formula = formula_va_ord_thres_1pl,
R>  data = VerbAgg,
R>  family = brmsfamily("cumulative", "logit"),
R>  prior = prior_va_ord_thres_1pl
R> )
\end{CodeInput}
\end{CodeChunk}

We can formally compare the two models using approximate LOO-CV which
reveals a LOOIC difference of \(\Delta \text{LOOIC}\) = -4.7 (SE = 14.1)
in favor of the simpler model with only shifted thresholds.
Acccordingly, we will stick to the assumption of shifted thresholds for
the remainder of this case study.

Similar to the binary case, one important extention to the standard GRM
is to assume varying discriminations across items. The resulting
generalized GRM is also a generalization of the binary 2PL model for
ordinal responses. We have seen in Section \ref{binary} that
discriminations were very similar across items in the binary case and I
now want to take a look again when modeling the ordinal responses. We
have to use slightly different formula syntax, though, as the non-linear
syntax of \pkg{brms} cannot handle the ordinal thresholds in the way
that is required when adding discrimination parameters. However, as
having discrimination parameters in ordinal models is crucial for IRT,
\pkg{brms} now provides a distributional parameter \texttt{disc}
specifically for that purpose. We can predict this discrimination
parameter using the distributional regression
framework\footnote{If \code{disc} is not predicted, it is automatically fixed to
\code{1}.}. By default, \texttt{disc} is modeled on the log-scale to
ensure that the actual discrimination estimates are positive (see
Section \ref{binary} for a discussion of that issue). The model formula
of the generalized GRM is given by

\begin{CodeChunk}

\begin{CodeInput}
R> formula_va_ord_2pl <- bf(
R+   resp ~ 1 + (1 |i| item) + (1 | id),
R+   disc ~ 1 + (1 |i| item)    
R+ )
\end{CodeInput}
\end{CodeChunk}

We specify some weakly informative priors on the hierarchical standard
deviations

\begin{CodeChunk}

\begin{CodeInput}
R> prior_va_ord_2pl <- 
R+   prior("constant(1)", class = "sd", group = "id") + 
R+   prior("normal(0, 3)", class = "sd", group = "item") +
R+   prior("normal(0, 1)", class = "sd", group = "item", dpar = "disc") 
\end{CodeInput}
\end{CodeChunk}

and finally fit the model:

\begin{CodeChunk}

\begin{CodeInput}
R> fit_va_ord_2pl <- brm(
R>   formula = formula_va_ord_2pl,
R>   data = VerbAgg,
R>   family = brmsfamily("cumulative", "logit"),
R>   prior = prior_va_ord_2pl
R> )
\end{CodeInput}
\end{CodeChunk}

A visualization of the item parameters can be found in Figure
\ref{fig:coef-item-va-ord-2pl}, in which we clearly see that
discrimination does not vary across items in the GRM either.

\begin{CodeChunk}
\begin{figure}

{\centering \includegraphics{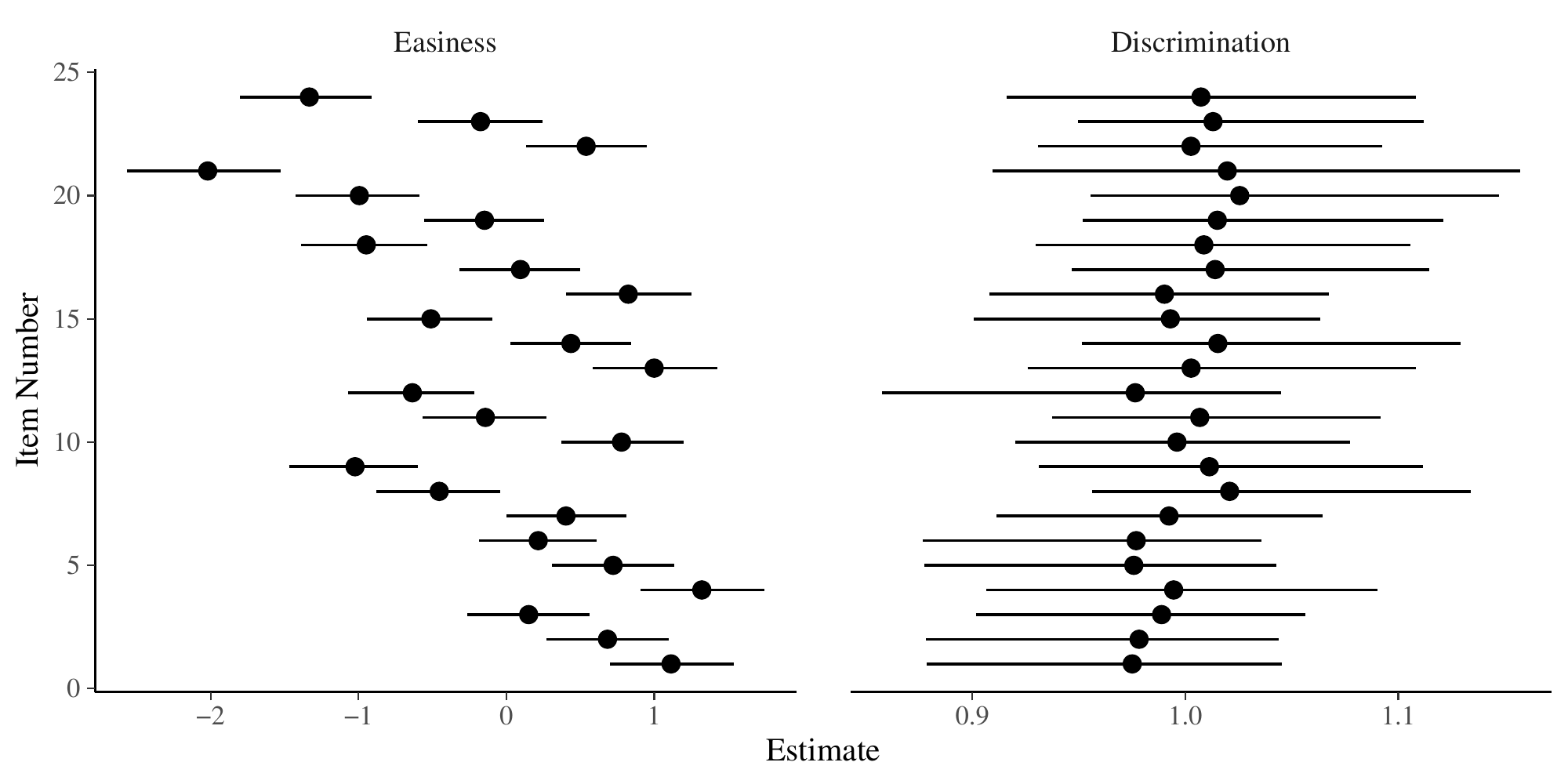} 

}

\caption[Posterior means and 95\% credible intervals of item parameters as estimated by model \code{fit\_va\_ord\_2pl}]{Posterior means and 95\% credible intervals of item parameters as estimated by model \code{fit\_va\_ord\_2pl}.}\label{fig:coef-item-va-ord-2pl}
\end{figure}
\end{CodeChunk}

Having made the decision to stick to the GRM with constant
discrimination, I again turn to the analysis of item and person
covariates. These can be specified in the same way as for binary models.
For instance, the GRM with both item and person covariates, and
interaction between \texttt{mode} and \texttt{Gender}, as well as
varying item parameters over \texttt{Gender} and varying person person
parameters over \texttt{mode} would look as follows:

\begin{CodeChunk}

\begin{CodeInput}
R> resp ~ Anger + Gender + btype + situ + mode + mode:Gender +
R>   (0 + Gender | item) + (0 + mode | id)
\end{CodeInput}
\end{CodeChunk}

We can fit the model as usual with the \texttt{brm} function and focus
on the effect of the trait \texttt{Anger} covariate in the following.
First, let us compare the regression coefficients of \texttt{Anger} as
obtained by the binary model and the GRM. We obtain \(b_{\rm 1PL} =\)
0.06 (\(95\% \; \text{CI}\) = {[}0.02, 0.09{]}) for the 1PL model and
\(b_{\rm GRM} =\) 0.07 (\(95\% \; \text{CI}\) = {[}0.04, 0.11{]}) for
the GRM, which are actually quite similar. Of course, this it not
necessarily true in general and we cannot know for sure before having
fitted both models. What will clearly be different are the predicted
response probabilities as we now have three instead of two categories:

\begin{CodeChunk}

\begin{CodeInput}
R> conditional_effects(fit_va_ord_cov1, effects = "Anger", categorical = TRUE)
\end{CodeInput}
\begin{figure}

{\centering \includegraphics{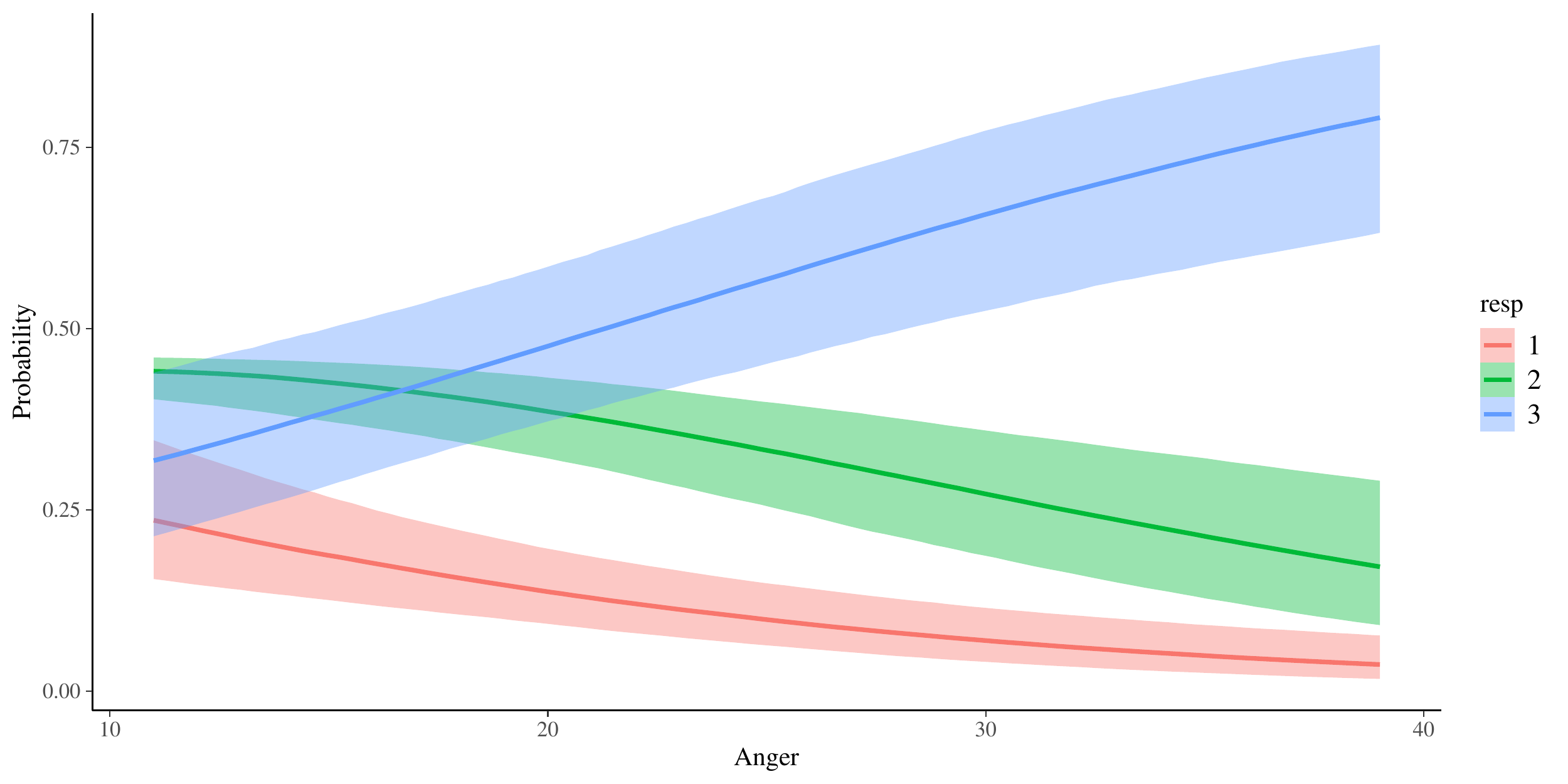} 

}

\caption[Expected probabilities of the three response categories in the \code{VerbAgg} data as a function of trait anger conditioned on all other categorical covariates being set to their reference categories and numerical covariates being set to their mean]{Expected probabilities of the three response categories in the \code{VerbAgg} data as a function of trait anger conditioned on all other categorical covariates being set to their reference categories and numerical covariates being set to their mean.}\label{fig:me-va-ord-cov1}
\end{figure}
\end{CodeChunk}

As can be seen in Figure \ref{fig:me-va-ord-cov1}, increased trait anger
is associated with higher probabilities of agreeing to items
(\texttt{yes}) as compared to choosing \texttt{no} or \texttt{perhaps}.
Although the plot may look like an interaction effect between
\texttt{Anger} and the response variable \texttt{resp}, it really is
just based on the single regression coefficient effecting the predicted
probabilities of all response categories. Plotting predicted response
probabilities instead of the response values themselves is recommend in
ordinal models as the latter assumes equidistant categories, which is
likely an invalid assumption for ordinal responses. That is, the
perceived difference between \texttt{no} and \texttt{perhaps} in the
participants' minds may be very different than the perceived difference
between \texttt{perhaps} and \texttt{yes}.

This is also what leads us to another potential problem with the model
assumptions, which is that the predictors are assumed to have a constant
effect across all response categories. For instance, it may very well be
that \texttt{Anger} has little effect on the choice between \texttt{no}
and \texttt{perhaps} but a much stronger one on the choice between
\texttt{perhaps} and \texttt{yes}. This can be explicitely modeled and
tested via what I call \emph{category specific effects}, which imply
estimating as many regression coefficients per category specific
predictor as there are thresholds (\(C - 1 = 2\) in our case).
Unfortunately, we cannot reliably model category specific effects in the
GRM as it may imply negative response category probabilities
\citep{buerkner2019}. Instead, we have to use another ordinal model and
I choose the \emph{partial credit model} \citep[PCM;][]{rasch1961} for
this purpose (see Section \ref{respdists} for details). In the PCM,
modeling category specific effects is possible because we assume not one
but \(C - 1\) latent variables which may have different predictor terms
(see Figure \ref{fig:pcm} for an illustration).

\begin{CodeChunk}
\begin{figure}

{\centering \includegraphics{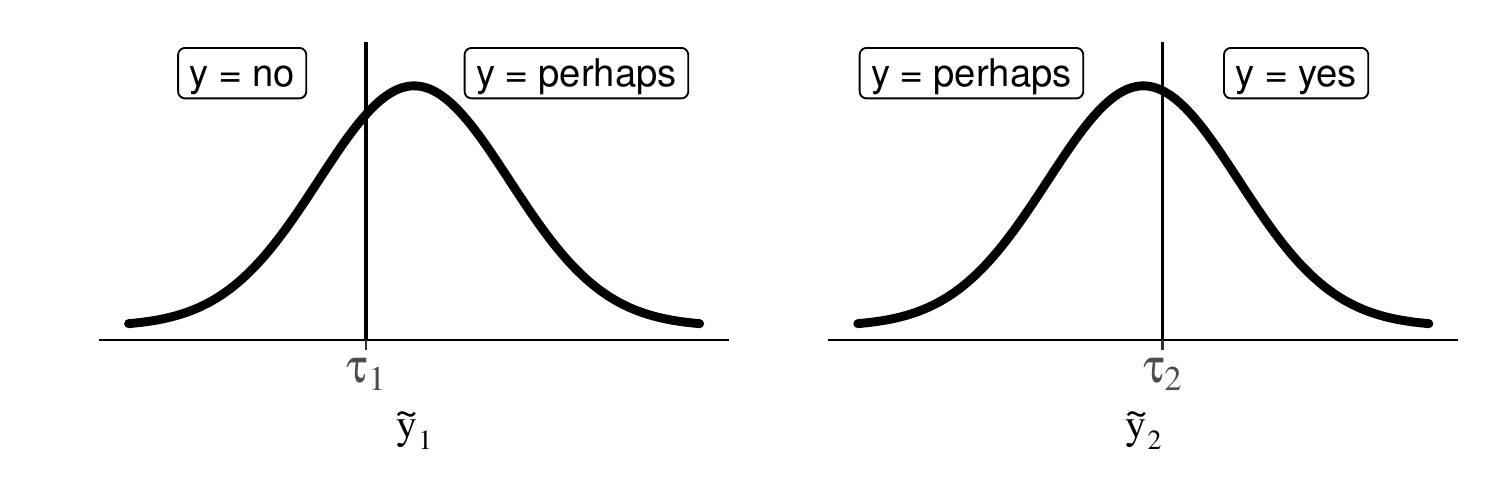} 

}

\caption{Assumptions of the partial credit model when applied to the \code{VerbAgg} data. The area under the curve in each bin represents the probability of the corresponding event given the set of possible events for the latent variables $\tilde{y}_1$ and $\tilde{y}_2$, respectively, which depend linearily on the predictor term $\eta$.}\label{fig:pcm}
\end{figure}
\end{CodeChunk}

Having selected an ordinal model class in which category specific
effects are possible, all we need to do is wrap the covariate in
\texttt{cs()} to estimate category specific effects. Suppose, we only
want to model \texttt{Anger} as category specific, then we replace
\texttt{Anger} with \texttt{cs(Anger)} in the model formula and leave
the rest of the formula unchanged:

\begin{CodeChunk}

\begin{CodeInput}
R> resp ~ cs(Anger) + Gender + btype + situ + mode + mode:Gender +
R>   (0 + Gender | item) + (0 + mode | id)
\end{CodeInput}
\end{CodeChunk}

The model is then fitted with \pkg{brms} in the same way as the GRM
except that we replace
\texttt{family\ =\ brmsfamily("cumulative",\ "logit")} by
\texttt{family\ =\ brmsfamily("acat",\ "logit")}. As the category
specific coefficients for \texttt{Anger} on the logit-scale we obtain
\(b_{\rm PCM 1}\) = 0.03 (\(95\% \; \text{CI}\) = {[}0, 0.05{]}) and
\(b_{\rm PCM 2}\) = 0.1 (\(95\% \; \text{CI}\) = {[}0.07, 0.13{]}). That
is, \texttt{Anger} seems to play a much stronger role in the decision
between \texttt{perhaps} and \texttt{yes} than between \texttt{no} and
\texttt{perhaps}. We may also visualize the effect via

\begin{CodeChunk}

\begin{CodeInput}
R> conditional_effects(fit_va_ord_cov2, effects = "Anger", categorical = TRUE)
\end{CodeInput}
\begin{figure}

{\centering \includegraphics{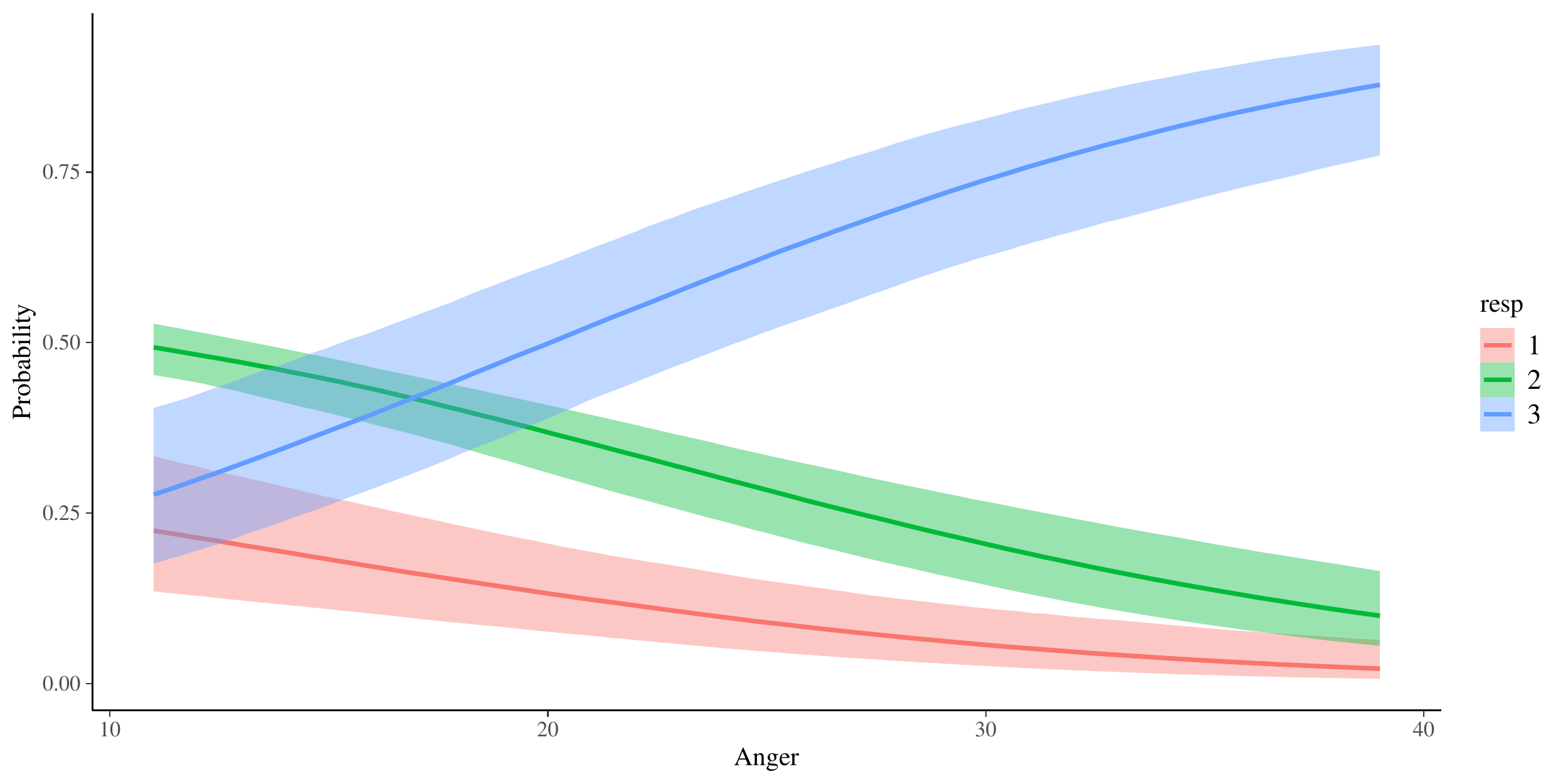} 

}

\caption[Expected response probabilities as predicted by model \code{fit\_va\_ord\_cov2} as a function of trait anger conditioned on all other categorical covariates being set to their reference categories and numerical covariates being set to their mean]{Expected response probabilities as predicted by model \code{fit\_va\_ord\_cov2} as a function of trait anger conditioned on all other categorical covariates being set to their reference categories and numerical covariates being set to their mean.}\label{fig:me-va-ord-cov2}
\end{figure}
\end{CodeChunk}

When we compare Figure \ref{fig:me-va-ord-cov2} to Figure
\ref{fig:me-va-ord-cov1}, we see that for higher \texttt{Anger} values a
higher probability of choosing \texttt{yes} and a lower probability of
choosing \texttt{perhaps} is predicted by the category specific PCM as
compared to the basic GRM. This is also in accordance with the
interpretation of the coefficients above.

\hypertarget{response-times}{%
\subsection{Response Times Models}\label{response-times}}

In this example, I will analyze a small data set of 121 subjects on 10
items measuring mental rotation that is shipped with the \pkg{diffIRT}
package \citetext{\citealp{diffIRT}; \citealp[see
also][]{vandermaas2011}}. The full data is described in
\citet{borst2011}. Each item consists of a graphical display of two
3-dimensional objects. The second object is either a rotated version of
the first one or a rotated version of a different object. The degree of
rotation (variable \texttt{rotate}) takes on values of 50, 100, or 150
and is constant for each item. Participants were asked whether the two
objects are the same (yes/no) and the response is stored as either
correct (1) or incorrect (0) (variable \code{resp}). The response time
in seconds (variable \code{time}) was recorded as well. A glimpse of the
data is provided in Table \ref{tab:head-rotation}.

\begin{CodeChunk}
\begin{table}

\caption{\label{tab:head-rotation}First ten rows of the \texttt{rotation} data.}
\centering
\begin{tabular}[t]{rrrrl}
\toprule
person & item & time & resp & rotate\\
\midrule
1 & 1 & 4.444 & 1 & 150\\
1 & 10 & 5.447 & 1 & 100\\
1 & 2 & 2.328 & 1 & 50\\
1 & 3 & 3.408 & 1 & 100\\
1 & 4 & 5.134 & 1 & 150\\
\addlinespace
1 & 5 & 2.653 & 1 & 50\\
1 & 6 & 2.607 & 1 & 100\\
1 & 7 & 3.126 & 1 & 150\\
1 & 8 & 2.869 & 1 & 50\\
1 & 9 & 3.271 & 1 & 150\\
\bottomrule
\end{tabular}
\end{table}

\end{CodeChunk}

I will start by analyzing the response times, only, and use the
exgaussian distribution for this purpose. Specifically, I am interested
in whether the degree of rotation affects the mean, variation and
right-skewness of the response times distribution. The effect of
\texttt{rotate} can be expected to be smooth and monotonic (up to 180
degrees after which the effect should be declining as the objects become
less rotated again) but otherwise of unknown functional form. In such a
case, it could be beneficial to model the effect via some
semi-parameteric methods such as splines or Gaussian processes (both of
which is possible in \pkg{brms}), but this requires considerable more
differentiated values of \texttt{rotate}. Thus, for this example, I will
just treat \texttt{rotate} as a factor and use dummy coding with \(50\)
degree as the reference category, instead of treating it as a continuous
variable. Assuming all three parameters to vary over persons and items,
we can write down the formula as

\begin{CodeChunk}

\begin{CodeInput}
R> bform_exg1 <- bf(
R+   time ~ rotate + (1 |p| person) + (1 |i| item),
R+   sigma ~ rotate + (1 |p| person) + (1 |i| item),
R+   beta ~ rotate + (1 |p| person) + (1 |i| item)
R+ )
\end{CodeInput}
\end{CodeChunk}

In theory, we could also model \texttt{rotate} as having a varying
effect across persons (as \texttt{rotate} is an item covariate).
However, as we are using a subset of only \(10\) items, modeling \(9\)
varying effects per person, although possible, will likely result in
overfitting. For larger data sets, this option could represent a viable
option and deserves further consideration. Since both \texttt{sigma}
(the standard deviation of the Gaussian component) and \texttt{beta}
(the mean parameter of the exponential component representing the right
skewness) can only take on positive values, I will use \texttt{log}
links for both of them (this is actually the default but I want to make
it explicit here). Together this results in the following model
specification:

\begin{CodeChunk}

\begin{CodeInput}
R> fit_exg1 <- brm(
R>   bform_exg1, data = rotation,
R>   family = brmsfamily("exgaussian", link_sigma = "log", link_beta = "log"),
R>   control = list(adapt_delta = 0.99)
R> )
\end{CodeInput}
\end{CodeChunk}

Increasing the sampling parameter \texttt{adapt\_delta} reduces or
ideally eliminates the number of ``divergent transition'' that indicate
problems of the sampler exploring the full posterior distribution and
thus bias the posterior estimates \citep{carpenter2017, hoffman2014}.
From the standard outputs (not shown here), we can see that the model
has converged well and produces reasonable posterior predictions (via
\code{pp_check(fit_exg1)}; see Figure \ref{fig:pp-exg1}), so we can turn
to investigating the effects of \texttt{rotate} on the model parameters:

\begin{CodeChunk}
\begin{figure}

{\centering \includegraphics{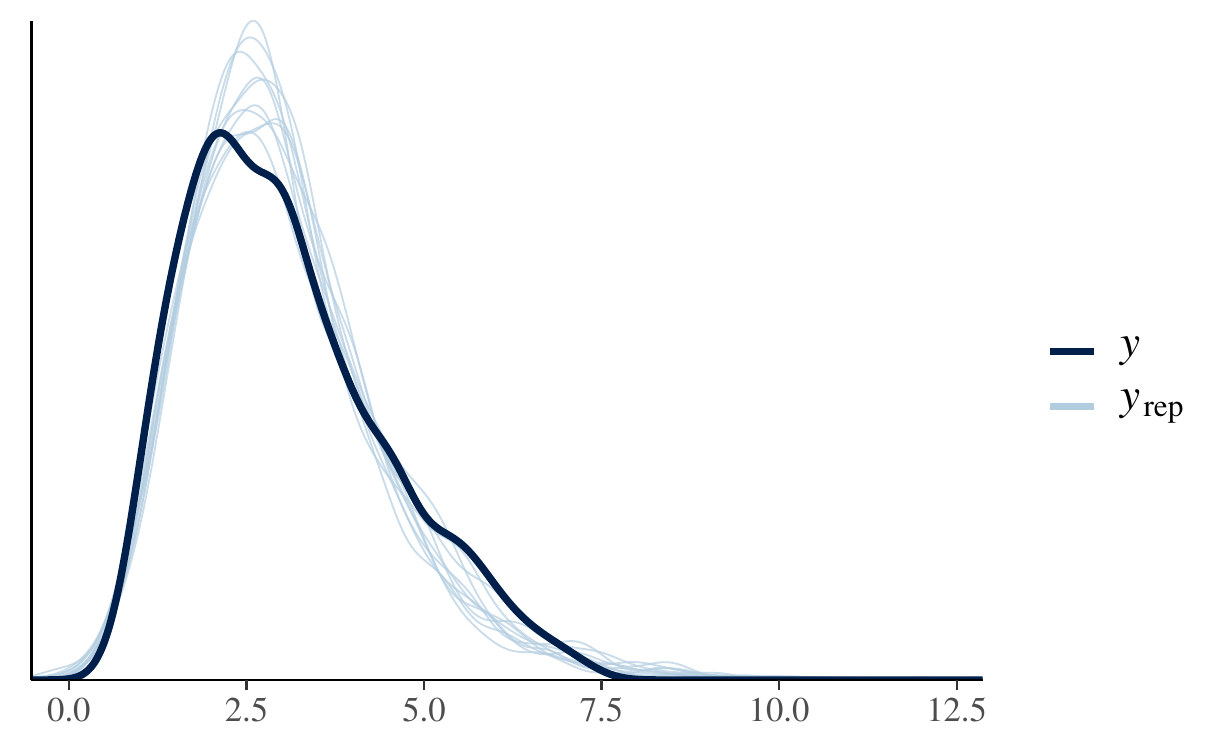} 

}

\caption[Posterior predictions of the exgaussian model \texttt{fit\_exg1}]{Posterior predictions of the exgaussian model \texttt{fit\_exg1}.}\label{fig:pp-exg1}
\end{figure}
\end{CodeChunk}

\begin{CodeChunk}

\begin{CodeInput}
R> conditional_effects(fit_exg1, "rotate", dpar = "mu")
R> conditional_effects(fit_exg1, "rotate", dpar = "sigma")
R> conditional_effects(fit_exg1, "rotate", dpar = "beta")
\end{CodeInput}
\end{CodeChunk}

\begin{CodeChunk}
\begin{figure}

{\centering \includegraphics{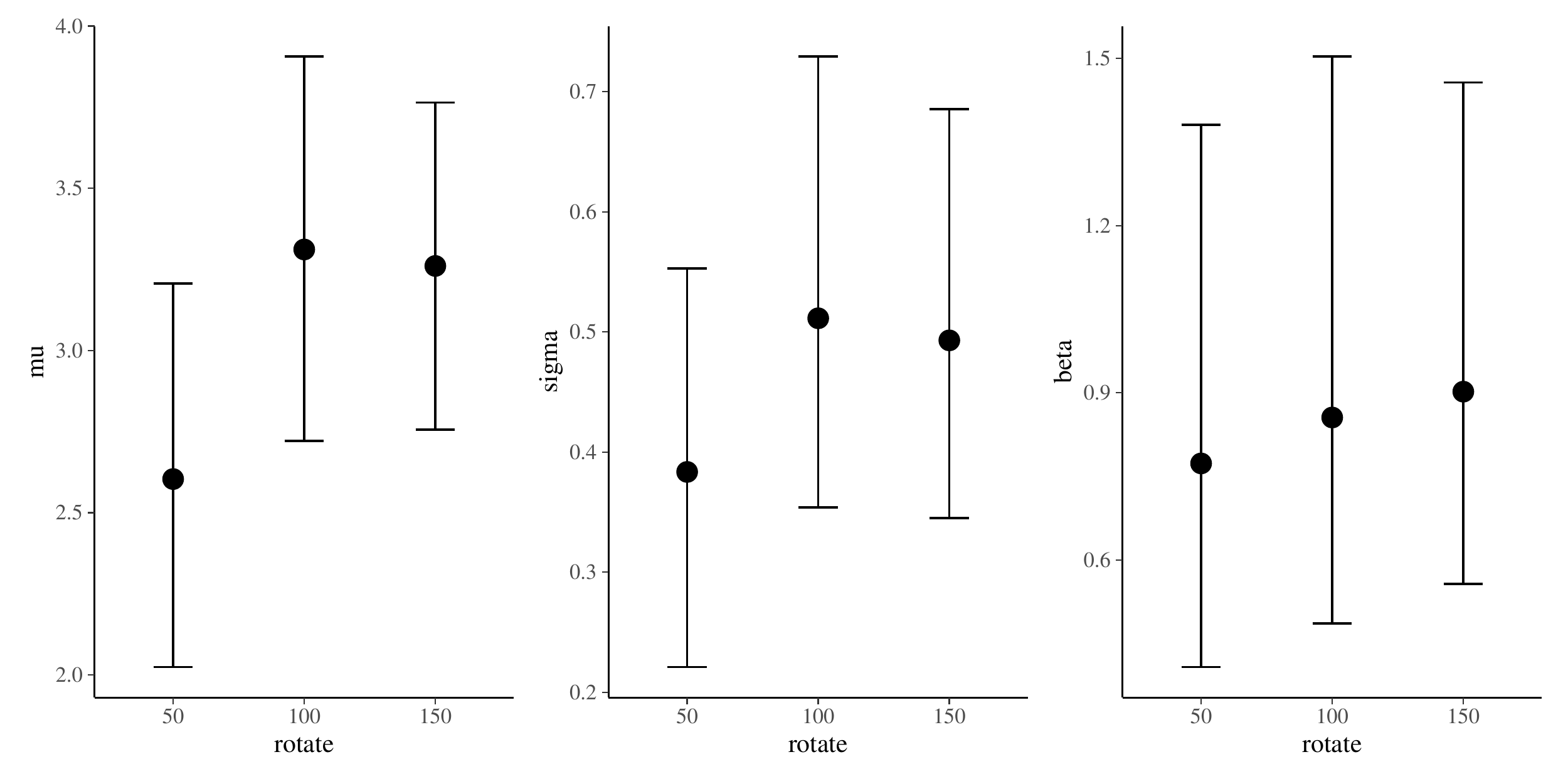} 

}

\caption[Parameters of the exgaussian model \texttt{fit\_exg1} as a function of the degree of rotation]{Parameters of the exgaussian model \texttt{fit\_exg1} as a function of the degree of rotation.}\label{fig:me-exg}
\end{figure}
\end{CodeChunk}

In Figure \ref{fig:me-exg}, we see that both the mean \texttt{mu} and
the variation \texttt{sigma} increase with increasing degree of
rotation, while the skewness \texttt{beta} roughly stays constant. The
observation that mean and variation of response times increase
simultaneously can be made in a lot of experiments and is discussed in
\citet{wagenmakers2007}.

The analysis of the response times is interesting, but does not provide
a lot of insights into potentially underlying cognitive processes. For
this reason, I will also use drift diffusion models to jointly model
response times and the binary decisions. How the drift diffusion model
looks exactly depends on several aspects. One is whether we deal with
personality or ability tests. For personality tests, the binary response
to be modeled is the actual \emph{choice} between the two alternatives,
whereas for ability tests may want to rather use the \emph{correctness}
instead \citep{tuerlinckx2005, vandermaas2011}. Further, in the former
case, person and item parameters may take on any real value and we
combine them additively. In contrast, for ability tests, person and item
parameters are assumed to be positive only and combined multiplicatively
\citep{vandermaas2011}. The latter can also be expressed as an additive
relationship on the log-scale. In the present example, we deal with data
of an ability test and will use the described log-scale approach.

Again, my interest lies primarily with the effect of the degree of
rotation. More specifically, I am interested in which of the three model
parameters (drift rate, boundary separation, and non-decision time) are
infludenced by the rotation. The fourth parameter, the initial bias, is
fixed to \(0.5\) (i.e., no bias) to obtain the three-parameter drift
diffuson model. Assuming all three predicted parameters to vary over
persons and items, we write down the formula as

\begin{CodeChunk}

\begin{CodeInput}
R> bform_drift1 <- bf(
R+   time | dec(resp) ~ rotate + (1 |p| person) + (1 |i| item),
R+   bs ~ rotate + (1 |p| person) + (1 |i| item),
R+   ndt ~ rotate + (1 |p| person) + (1 |i| item),
R+   bias = 0.5
R+ )
\end{CodeInput}
\end{CodeChunk}

In \proglang{Stan}, drift diffusion models with predicted non-decision
time are not only computationally much more demanding, but they also
often require some manual specifcation of initial values. The easiest
way is to set the intercept on the log-scale of \texttt{ndt} to a small
value:

\begin{CodeChunk}

\begin{CodeInput}
R> chains <- 4
R> inits_drift <- list(temp_ndt_Intercept = -3)
R> inits_drift <- replicate(chains, inits_drift, simplify = FALSE)
\end{CodeInput}
\end{CodeChunk}

I will now fit the model. This may take some more time than previous
models due to the complexity of the diffusion model's likelihood.

\begin{CodeChunk}

\begin{CodeInput}
R> fit_drift1 <- brm(
R>   bform_drift, data = rotation, 
R>   family = brmsfamily("wiener", "log", link_bs = "log", link_ndt = "log"),
R>   chains = chains, cores = chains,
R>   inits = inits_drift, init_r = 0.05,
R>   control = list(adapt_delta = 0.99)
R> )
\end{CodeInput}
\end{CodeChunk}

From the standard outputs (not shown here), we can see that the model
has converged well so we can turn to investigating the effects of
\texttt{rotate} on the model parameters:

\begin{CodeChunk}

\begin{CodeInput}
R> conditional_effects(fit_drift1, "rotate", dpar = "mu")
R> conditional_effects(fit_drift1, "rotate", dpar = "bs")
R> conditional_effects(fit_drift1, "rotate", dpar = "ndt")
\end{CodeInput}
\end{CodeChunk}

\begin{CodeChunk}
\begin{figure}

{\centering \includegraphics{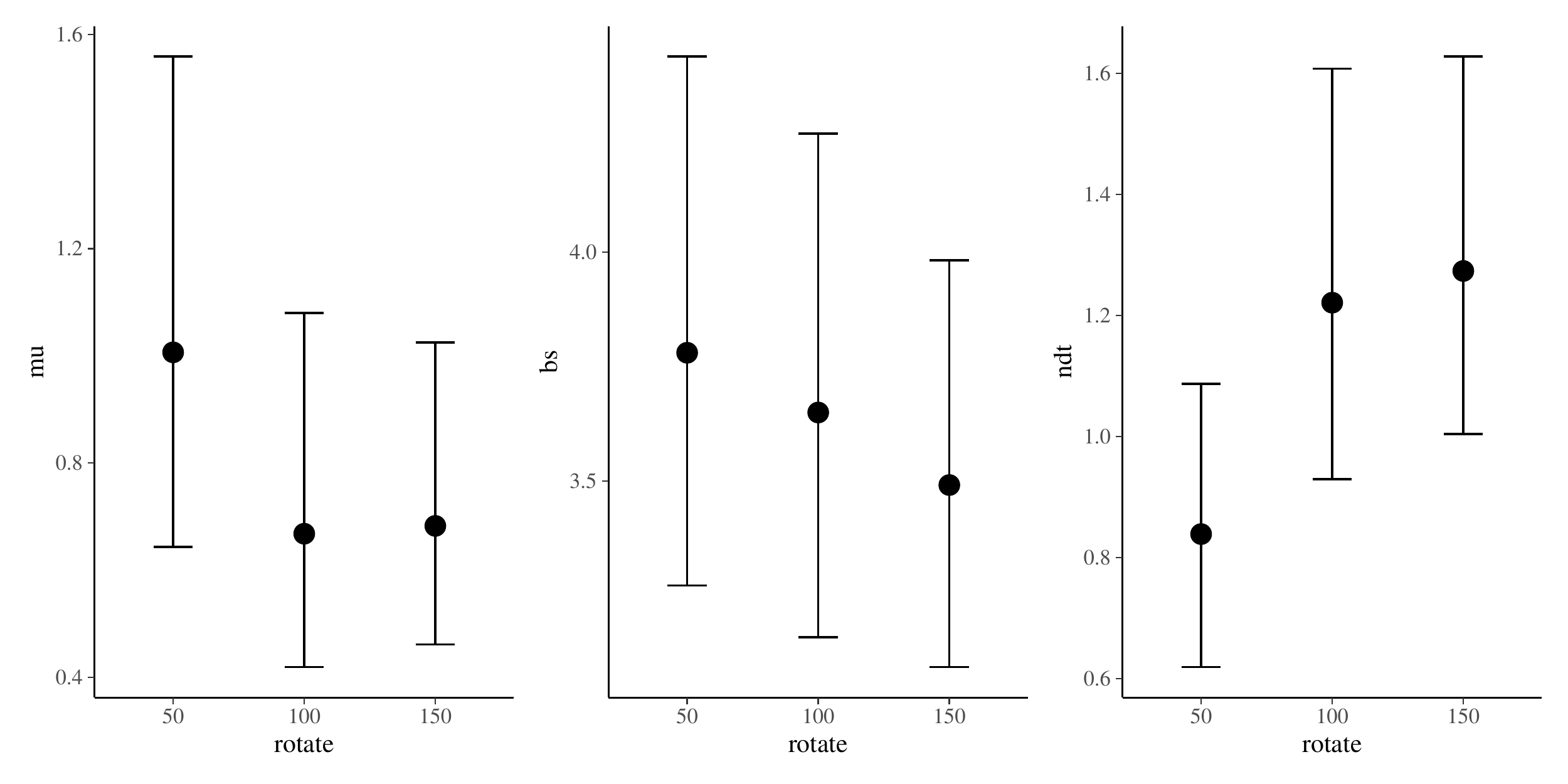} 

}

\caption[Parameters of the drift diffusion models as a function of the degree of rotation]{Parameters of the drift diffusion models as a function of the degree of rotation. The parameter \texttt{mu}, \texttt{bs}, and \texttt{ndf} represent the drift rate, boundary separation and non-decision time, respectively.}\label{fig:me-drift}
\end{figure}
\end{CodeChunk}

As shown in Figure \ref{fig:me-drift}, both the drift rate and the
non-decision time seem to be affected by the degree of rotation. The
drift rate decreases slightly when increasing the rotation from \(50\)
to \(100\) and roughly stays constant afterwards. Similarily, the
non-decision time increases with increased rotation from \(50\) to
\(100\) presumably as a result of the increased cognitive demand of
processing the rotated objects \citep{diffIRT}.

In contrast, the boundary separation appears to be unaffected by the
degree of rotation. Further, the standard deviation of the boundary
separation across items (after controlling for the rotation), seems to
be very small (SD = 0.05, 95\%-CI = {[}0, 0.16{]}). We may also test
this more formally by fitting a second model without item effects on the
boundary separation, that is using the formula
\texttt{bs\ \textasciitilde{}\ 1\ +\ (1\ \textbar{}p\textbar{}\ person)},
and then comparing the models for instance via approximate LOO-CV
(method \texttt{loo}) or Bayes factors (method \texttt{bayes\_factor}).
The latter requires carefully specified prior distributions based on
subject matter knowledge, a topic which is out of the scope of the
present paper.

\hypertarget{comparison}{%
\section{Comparison of Packages}\label{comparison}}

A lot of \proglang{R} packages have been developed that implement IRT
models, each being more or less general in their supported models. In
fact, for most IRT models developed in the statistical literature, we
may actually find an \proglang{R} package implementing it. An overview
of most of these packages is available on the Psychometrics CRAN task
view (\url{https://cran.r-project.org/web/views/Psychometrics.html}).
Comparing all of them to \pkg{brms} would be too extensive and barely
helpful for the purpose of the present paper. Accordingly, I focus on a
set of nine widely applied and actively maintained packages that can be
used for IRT modeling. These are \pkg{eRm} \citep{eRm}, \pkg{ltm}
\citep{ltm}, \pkg{TAM} \citep{TAM}, \pkg{mirt} \citep{mirt}, \pkg{sirt}
\citep{sirt}, \pkg{lme4} \citep{lme4}, \pkg{lavaan} \citep{lavaan},
\pkg{blavaan} \citep{blavaan}, and \pkg{MCMCglmm} \citep{MCMCglmm}. All
of these packages are of high quality, user friendly, and well
documented so I primarily focus my comparison on the features they
support. A high level overview of the modeling options of each package
can be found in Table \ref{tab:pkg-compare} and more details are
provided below.

\pkg{eRm} focuses on models that can be estimated using conditional
maximum likelihood, a method only available for the 1PL model and PCM
with unidimensional latent traits per person. Accordingly, its
application is the most specialized among all the packages presented
here. The \pkg{ltm}, \pkg{TAM}, and \pkg{mirt} packages all provide
frameworks for fitting binary, categorical, and ordinal models using
mostly marginal maximum likelihood estimation. They allow estimating
discrimination parameters for all of these model classes as well as 3PL
or even 4PL models for binary responses. Of these three packages,
\pkg{mirt} currently provides the most flexible framework with respect
to both the models it can fit and the provided estimation algorithms.
The package also comes with its own modeling syntax for easy
specification of factor structure and parameter constraints and
implements exploratoy multilevel IRT models via the \code{mixedmirt}
function. The \pkg{sirt} package, does not provide one single framework
for IRT models but rather a large set of separate functions to fit
special IRT models that complement and support other packages, in
particular \pkg{mirt} and \pkg{TAM}. As a result, input and output
structures are not consistent across model fitting functions within
\pkg{sirt}, which makes it more complicated to switch between model
classes. All of these IRT-specific packages have built-in methods for
investigating and testing differential item functioning. In addition to
these tools, a powerful approach for assessing differential item
functioning via recursive partioning is implemented in \pkg{psychotree}
\citep{psychotree1, psychotree2} based on methods of the
\pkg{psychotools} package \citep{psychotools}. It currently supports
methods for dichotomous, categorical, and ordinal models.

In contrast to the above packages, \pkg{lavaan} and \pkg{lme4} are not
specifically dedicated to IRT modeling, but rather provide general
frameworks for structural equation and multilevel models, respectively.
Due to their generality and user-friendly interfaces, they have
established themselves as the de facto standards in \proglang{R} when it
comes to the frequentist estimation of these model classes. \pkg{lavaan}
allows to fit multidimensional 1PL and 2PL binary, categorical and some
ordinal IRT models using maximum likelihood or weighted least squares
estimations. In addition, the \pkg{blavaan} package allows to fit
\pkg{lavaan} models using Bayesian estimation methods. To date, not all
\pkg{lavaan} models are available in \pkg{blavaan}, but I expect this to
change in the future. \pkg{lme4} estimates multilevel models via
marginal maximum likelihood estimation. While it is very flexible in the
specification of covariates and multilevel structure, for instance, for
the purpose of multidimensional IRT models, it neither supports 2PL (or
more parameters) binary models, nor categorical, or ordinal models.
\pkg{MCMCglmm} implements multivariate multilevel models in a Bayesian
context and provides a wider range of model families but, similar to
\pkg{lme4}, has limitations in the area of non-linear IRT models such as
2PL (or more parameters) binary models.

\pkg{brms} is conceptually closest to \pkg{lme4} and \pkg{MCMCglmm} when
it comes to the model specification and data structuring. These three
packages expect the data to be in long format, that is all responses to
be provided in the same column, while all other packages expect response
to be in the form of a person \(\times\) item matrix. Accordingly, the
formula syntax also differs from the other packages in that we have to
explicitely specify item and person grouping variables as they cannot be
automatically identified from the data structure (see Section
\ref{brms}). The multilevel syntax of \pkg{lme4}, \pkg{MCMCglmm}, and
\pkg{brms} allows for an overall shorter model specification than the
structural equation syntax of \pkg{lavaan} as items do not have to be
targeted one by one. A drawback of the multilevel syntax is that
constraining or fixing parameters is perhaps less intuitive than in the
dedicated IRT packages or \pkg{lavaan} syntax.

What makes \pkg{brms} stand out is the combination of three key
features. First, it extends the multilevel formula syntax of \pkg{lme4}
to non-linear formulas of arbitrary complexity, which turns out to be
very powerful for the purpose of IRT modeling (see Section \ref{brms}).
Second, it supports the widest range of response distributions of all
the packages under comparison. This includes not only distributions for
binary, categorical, and ordinal data, but also for response times,
count, or even proportions to name only a few available options.
Further, users may specify their own response distributions via the
\code{custom_family} feature, fulfilling a similar purpose as
\code{mirt::createItem} or \code{sirt::xxirt}. Third, not only the main
location parameter but also all other parameters of the response
distribution may be predicted by means of the non-linear multilevel
syntax. In addition, multiple different response variables can be
combined into a joint multivariate model in order to let person and/or
item parameters inform each other, respectively, across response
variables.

Another difference between \pkg{brms} and several of the other packages
is that the former is fully Bayesian while the latter are mostly based
on point estimation methods. \pkg{TAM} and \pkg{mirt} support setting
certain prior distributions on parameters but still perform estimation
via optimization. \pkg{sirt} offers MCMC sampling only for 2PL and 3PL
models with restrictive prior options and few built-in methods to
post-process results. Similarily, \pkg{blavaan} can fit a subset of the
models supported by \pkg{lavaan} using MCMC methods implemented in
\proglang{JAGS} \citep{jags} or \proglang{Stan} \citep{carpenter2017}
although the set of supported IRT models is currently much smaller than
that of \pkg{brms} (see Table \ref{tab:pkg-compare}). \pkg{MCMCglmm}
also provides MCMC estimation for a quite wide range of families but
also falls short of the flexibility of brms, for example, in the context
of non-linear models or response times distributions. While performing
full Bayesian inference via MCMC sampling is often orders of magnitude
slower than point estimation via maximum likelihood or least squares,
the obtained information may be considered to be much higher: Not only
do we get the posterior distribution of all model parameters, but also
the posterior distribution of all quantities that can be computed on
their basis \citep{gelman2013}. For instance, the uncertainty in the
parameters' posterior naturally propagates to the posterior predictive
distributions, whose uncertainty can then be visualized along with the
mean predictions. \pkg{brms} automates a lot of common post-processing
tasks, such as posterior visualizations, predictions, and model
comparison (see \code{methods(class =
"brmsfit")} for a full list of options and the replication material of
this paper for examples).

To what extent the increased information obtained via full Bayesian
inference is worth the additional computational costs and corresponding
waiting time depends on various factors related to the model, data, and
goal of inference. For instance, if the model is relatively simple and
there is a lot of data available to inform the model parameters,
Bayesian and maximum likelihood estimates are unlikely to differ a lot
unless strong prior information is provided. Also, if the goal is to
provide estimates in real time, for instance for the purpose of adaptive
testing, full Bayesian inference may be too slow to be viable unless
specifically tuned to such a task \citep[e.g., see][]{vanderlinen2015}.
I do not argue that a Bayesian approach to IRT is always superior, but
instead want to point out its strengths (and also its weaknesses) so
that users can make an informed decision as to when working with a
Bayesian framework may improve the desired inference \citep[see also][
for an example in the context of IRT]{buerkner2020spm}.

Similarily, while using general purpose frameworks for IRT such as those
provided by \pkg{brms}, \pkg{lme4}, \pkg{lavaan}, or \pkg{MCMCglmm} may
provide advantages in terms of modeling flexibility and consistency of
model specification and post-processing, they clearly come with some
disadvantages. Among others, such general frameworks are likely to
require more work from the user at the start to familiarize themselves
with the interface in order to fit the desired models as compared to
packages with specific built-in function for common model classes. At
the same time, post-processing methods of specialized software may be
easier and more directly applicable to common use-cases, thus lowering
the requirements in the actual coding expertise of users. For instance,
the specification and post-processing of standard 1PL or 2PL models is
more straightforward in dedicated IRT software and users only interested
in such models may get reliable solutions faster this way. In other
words, when introducing more and more general frameworks, the goal is
not to render more specialized software irrelevant, but to provide an
alternative for consistent model building and evaluation with a larger
scope than specialized software is intended for.

\begin{CodeChunk}
\begin{table}

\caption{\label{tab:pkg-compare}Overview of modeling options in some IRT supporting packages.}
\centering
\begin{threeparttable}
\resizebox{\linewidth}{!}{
\begin{tabular}[t]{llllllllll}
\toprule
\multicolumn{1}{c}{ } & \multicolumn{9}{c}{Package} \\
\cmidrule(l{3pt}r{3pt}){2-10}
Feature & eRm & ltm & TAM & mirt & sirt & (b)lavaan & lme4 & MCMCglmm & brms\\
\hline
1-PLM & yes & yes & yes & yes & yes & yes & yes & yes & yes\\
2-PLM & no & yes & yes & yes & yes & yes & no & no & yes\\
3-PLM & no & yes & yes & yes & yes & no & no & no & yes\\
4-PLM & no & no & no & yes & yes & no & no & no & yes\\
\addlinespace
PCM & yes & yes & yes & yes & yes & no & no & no & yes\\
GRM & no & yes & no & yes & yes & yes & no & yes & yes\\
CM & no & no & yes & yes & no & no & no & yes & yes\\
LM & no & no & no & no & yes & yes & yes & yes & yes\\
CoM & no & no & no & no & no & no & yes & yes & yes\\
\addlinespace
RTM & no & no & no & no & no & limited & limited & limited & yes\\
PrM & no & no & no & no & no & no & no & no & yes\\
\hline
Multidimensional & no & no & yes & yes & yes & yes & yes & yes & yes\\
Covariates & yes & yes & yes & yes & yes & yes & yes & yes & yes\\
Constraints & no & yes & yes & yes & yes & yes & limited & limited & limited\\
\addlinespace
Latent classes & no & no & yes & yes & yes & no & no & no & no\\
Mixtures & no & no & yes & yes & yes & no & no & no & yes\\
Copulas & no & no & limited & no & limited & no & no & no & no\\
Splines & no & no & no & yes & yes & no & no & no & yes\\
Multilevel & no & no & no & yes & limited & limited & yes & yes & yes\\
\addlinespace
Joint models & no & no & no & yes & no & yes & no & yes & yes\\
Imputation & no & no & yes & yes & yes & no & no & no & yes\\
Customizable & no & no & no & yes & yes & no & no & no & yes\\
Estimator & CML & MML & MML,JML & MML & various & various & MML & MH-Gibbs & AHMC\\
\bottomrule
\end{tabular}}
\begin{tablenotes}
\item \small Abbreviations: x-PLM = x-parameter logistic models; PCM = partial credits models; GRM = \newline graded response models; CM = categorical models; LM = linear models; CoM = count data \newline  models; RTM = response times models; PrM =  Proportion models (i.e., Beta and Dirichlet models); \newline  CML = conditional maximum likelihood; MML = marginal maximum likelihood; JML = joint \newline maximum likelihood; MH-Gibbs = Metropolis-Hastings within Gibbs Sampler; AHMC = adaptive \newline Hamiltonian Monte-Carlo.
\end{tablenotes}
\end{threeparttable}
\end{table}

\end{CodeChunk}

\hypertarget{conclusion}{%
\section{Conclusion}\label{conclusion}}

In this paper, I have introduced a general framework for fitting
Bayesian IRT models in \proglang{R} via \pkg{brms} and \proglang{Stan}.
Within this framework, a wide range of IRT models can be specified,
estimated, and post-processed in a consistent manner, without the need
to switch between packages to obtain results for different IRT model
classes. I have demonstrated its usefulness in examples of binary,
ordinal, and response times data, although the framework entails a lot
of other IRT model classes.

The advanced formula syntax of \pkg{brms} further enables the modeling
of complex non-linear relationships between person and item parameters
and the observed responses. However, the flexibility of the framework
does not free the user from specifying reasonable models for their data.
Just because a model can be estimated without problems does not mean it
is also sensible from a theoretical perspective or provides valid
inference about the effects under study. Tools for model comparison and
selection as provided by \pkg{brms} may help in guiding users' decision,
but should not be a substitute for clear theoretical reasoning and
subject matter knowledge to guide model development and evaluation.

Taking a Bayesian perspective on specification, estimation, and
post-processing of statistical models helps in building and fitting more
complex and realistic models, but it is not the only reason for adopting
it. As Bayesian statistics is fully embedded into probability theory, we
can quantify uncertainty of any variable of interest using probability
and make decisions by averaging over that uncertainty. Thus, we no
longer have to fall back on premature binary decision making on the
basis of, say, frequentist p-values or confidence intervals. As such,
Bayesian inference is not just another estimation method but a distinct
statistical framework to reason from data using probabilistic models.

\hypertarget{acknowledgements}{%
\section{Acknowledgements}\label{acknowledgements}}

I would like to thank my colleagues of the \proglang{Stan} Development
Team for creating, maintaining, and continuously improving
\proglang{Stan}, which forms the basis for the success of \pkg{brms}.
Further, I want to thank Marie Beisemann, Alexander Robitzsch, and two
anonymous reviewers for valuable comments on earlier versions of the
paper. Finally, I would like to thank all the users who reported bugs or
had ideas for new features, thus helping to further improve \pkg{brms}.

\renewcommand\refname{References}
\bibliography{Bayesian-IRT}

\end{document}